\documentclass[superscriptaddress, pra, twocolumn, amsmath, amssymb, aps]{revtex4-2}

\usepackage{graphicx}% Include figure files
\usepackage{dcolumn}% Align table columns on decimal point
\usepackage{bm}% bold math
\usepackage[dvipsnames]{xcolor}
\usepackage{graphicx}
\usepackage{amsmath}

\usepackage{tikz}
\usetikzlibrary{optics}
\usetikzlibrary{shapes,arrows}
\usetikzlibrary {chains}
\usepackage{siunitx}
\usepackage{quantikz}
\usepackage{braket}
\usepackage{adjustbox}
\usepackage{stackrel}
\usepackage{comment}

\usepackage{algorithm}
\usepackage{algpseudocode}

\usepackage[hidelinks]{hyperref}
\usepackage{orcidlink}

\hypersetup{
    unicode=false,          % non-Latin characters in AcrobatÕs bookmarks
    pdftoolbar=false,        % show AcrobatÕs toolbar?
    pdfmenubar=true,        % show AcrobatÕs menu?
    pdffitwindow=false,     % window fit to page when opened
    pdfstartview={FitH},    % fits the width of the page to the window
    pdftitle={},    % title
    pdfcreator={},   % creator of the document
    pdfproducer={}, % producer of the document
    pdfkeywords={error detection, quantum state tomography, quantum optics}, % list of keywords
    pdfnewwindow=true,      % links in new window
    colorlinks=true,       % false: boxed links; true: colored links
    linkcolor=blue,          % color of internal links (change box color with linkbordercolor)
    citecolor=blue,        % color of links to bibliography
    filecolor=teal,      % color of file links
    urlcolor=blue           % color of external links
}

\def\rhoUnbiased{{\rho_{\text{U}}}}  
\def\rhoBiased{{\rho_{\text{B}}}}  
\def\rhoLin{{\rho_{\text{LIN}}}}  
\def\rhoLS{{\rho_{\text{LS}}}}

\def\rhoNonPhys{{\hat{\rho}_{\text{NP}}}}

\def\PauliX{{\sigma_{\text{X}}}} 
\def\PauliY{{\sigma_{\text{Y}}}} 
\def\PauliZ{{\sigma_{\text{Z}}}} 
\def\trace{{\operatorname{Tr}}} 

\DeclareMathOperator*{\argmax}{arg\,max}
\DeclareMathOperator*{\argmin}{arg\,min}
\usepackage{xspace}
\newcommand{\ea}{\textit{et al.}\xspace}

\begin{document}

\preprint{APS/123-QED}

\title{Ability of entanglement and purity to help to detect systematic experimental errors}

\author{Julia Freund\orcidlink{0000-0001-5548-5007}}
\affiliation{Institut für Theoretische Physik, Universität Innsbruck, Technikerstraße 21a, 6020 Innsbruck, Austria}
\affiliation{Institute of Semiconductor and Solid State Physics, Johannes Kepler University, Altenberger Straße 69, 4040 Linz, Austria}

\author{Francesco Basso Basset\orcidlink{0000-0002-2148-0010}}
 \affiliation{Department of Physics, Sapienza University of Rome, Piazzale Aldo Moro 5, 00185 Rome, Italy}
\affiliation{Dipartimento di Fisica, Politecnico di Milano, Piazza Leonardo da Vinci 32, 20133 Milano, Italy}

\author{Tobias M. Krieger}
\affiliation{Institute of Semiconductor and Solid State Physics, Johannes Kepler University, Altenberger Straße 69, 4040 Linz, Austria}

\author{Alessandro Laneve}
 \affiliation{Department of Physics, Sapienza University of Rome, Piazzale Aldo Moro 5, 00185 Rome, Italy}

\author{Mattia Beccaceci} 
 \affiliation{Department of Physics, Sapienza University of Rome, Piazzale Aldo Moro 5, 00185 Rome, Italy}

\author{Michele B. Rota\orcidlink{0000-0003-3933-6997}} 
 \affiliation{Department of Physics, Sapienza University of Rome, Piazzale Aldo Moro 5, 00185 Rome, Italy}

\author{Quirin Buchinger}
\affiliation{Technical Physics, University of Wuerzburg, Am Hubland, 97074 Wuerzburg, Germany}

\author{Saimon F. Covre da Silva}
\affiliation{Institute of Semiconductor and Solid State Physics, Johannes Kepler University, Altenberger Straße 69, 4040 Linz, Austria}
\affiliation{Instituto de Física Gleb Wataghin, Universidade Estadual de Campinas (UNICAMP), 13083-859 Campinas, Brazil}

\author{Sandra Stroj}
\affiliation{Forschungszentrum Mikrotechnik, FH Vorarlberg, Hochschulstraße 1, 6850 Dornbirn, Austria}

\author{Sven Höfling}
\affiliation{Technical Physics, University of Wuerzburg, Am Hubland, 97074 Wuerzburg, Germany}

\author{Tobias Huber-Loyola}
\affiliation{Technical Physics, University of Wuerzburg, Am Hubland, 97074 Wuerzburg, Germany}
\affiliation{Institute of Photonics and Quantum Electronics \& IQST, Karlsruhe Institute of Technology, Engesserstr. 5, 76131 Karlsruhe, Germany}

\author{Richard Kueng}
\affiliation{Institute for Integrated Circuits, Johannes Kepler University, Altenberger Straße 69, 4040 Linz, Austria}

\author{Armando Rastelli} 
 \affiliation{Institute of Semiconductor and Solid State Physics, Johannes Kepler University, Altenberger Straße 69, 4040 Linz, Austria}

 \author{Rinaldo Trotta\orcidlink{0000-0002-9515-6790}} 
 \affiliation{Department of Physics, Sapienza University of Rome, Piazzale Aldo Moro 5, 00185 Rome, Italy}

 \author{Otfried Gühne\orcidlink{0000-0002-6033-0867}}
  \affiliation{Naturwissenschaftlich-Technische Fakultät, Universität Siegen, Walter-Flex-Str. 3, 57068 Siegen, Germany}

\date{\today}

\begin{abstract}
Measurements are central in all quantitative sciences, and a fundamental challenge is to make observations without systematic measurement errors. This holds in particular for quantum information processing, where other error sources, such as noise and decoherence, are unavoidable. Consequently, methods for detecting systematic errors have been developed, but the required quantum state properties are yet unexplored. We theoretically develop a direct and efficient method to detect systematic errors in quantum experiments and demonstrate it experimentally using quantum state tomography of photon pairs emitted from a semiconductor quantum dot. Our method can be scaled to multi-qubit systems, and we find that entanglement and quantum states with high purity can help identify systematic errors. 
\end{abstract}

\maketitle

\section{Introduction}
The last 30 years in quantum information science~\cite{nielsen_chuang_2010, HorodeckiActaPhysicaPolonicaA, Zoller2005} promise exciting applications such as quantum communication~\cite{Chen2021ReviewCommComp,QuantumInternetWehner, Foreman2023practicalrandomness}, efficient problem-solving beyond the reach of classical computers~\cite{alex2023quantumalgo, Harrow2017}, and the simulation of complex many-body systems.~\cite{Fauseweh2024, BulutaScienceQSimulator2009, Houck2012, Aspuru-Guzik2012}. All of these applications strongly rely on the correct readout of quantum information via measurements of a quantum system. For instance, quantum state tomography \cite{paris2004quantum} and shadow tomography~\cite{Huang2020,Nguyen2022PRL} are typical examples of an information readout procedure that utilizes a finite number of measurements of equally prepared state copies. In state tomography, however, measurement errors can lead to nonphysical estimates of the quantum state. In practice, statistical and systematic errors are the most prominent types 
of measurement errors, and they are relevant not only in quantum state tomography but also in other tools to analyze quantum systems~\cite{MoroderRobustEntDetPRA, TavakoliPRL2021, Cao2023GenuineMultipartiteEnt, Svegborn2024imprecision, Cieslinski2024,Woelk_2019}.

Statistical errors in measurements arise from finite statistics: each experiment is repeated only a finite number of times and the observed frequencies of outcomes do not necessarily correspond to the outcome probabilities. In quantum state tomography this can lead to nonphysical state estimates. The current literature addresses statistical errors by discussing confidence regions of estimators~\cite{relieableStateTomographyChristandlRenner, Sugiyama_2012, PolytopesRenner, errorBarsRenner, Guta_2020, degois2023userfriendly} or by using statistical tools~\cite{Yu_2022} to find the necessary number of measurement repetitions to recover a physical state estimate~\cite{knips2015long}. 

In contrast to statistical errors, systematic errors originate from various environmental influences and imperfections, and various works discussed the impact of errors like misalignment of the measurement basis. This concerns determining the state fidelity and entanglement witnesses~\cite{Gisin2012PRA}, systematic errors due to bias in quantum state
estimators~\cite{SchwemmerPRL2015, SilvaPRA2017}, robustness of tomography schemes~\cite{OptimalSchemes,WangPRA2023,Ivanova-Rohling2023} and the characterization of photon detectors~\cite{MeasuringMeasurementsFeito,Novik2019}. All of these results emphasize the variety and relevance of statistical and systematic errors in the field of quantum information. 

This naturally leads to the question of how to distinguish statistical from systematic errors. Both error types can have similar effects, but while repeated experiments can suppress statistical errors, systematic errors may reveal fundamental flaws in the experiment. Ref.~\cite{witnessesPRL} constructed witnesses to certify systematic errors from measurement data. Moreover, Ref.~\cite{Langford_2013} discussed the chi-squared goodness-of-fit test to assess the quality of the reconstructed state with respect to a previously chosen model and how to modify the test appropriately for states close to the border of the physical state space.

In this work, we first develop theoretically a direct and efficient method to detect systematic errors. Then, we experimentally implement the scheme using entangled photon pairs emitted by a semiconductor quantum dot \cite{SchimpfPerspective2021}. We employ strain tuning on the source~\cite{rota2024source,trottaPRl2015} to generate two-photon polarization states with a varying degree of entanglement and purity. We use quantum state tomography as an example for an involved quantum information task, but one can adjust the ideas for other quantum tasks and experimental platforms as well. Our findings demonstrate that even quantum states with a low purity can be sensitive enough to signal the presence of systematic errors. Moreover, if two or more particles are considered, entanglement of the probe states can be essential in our scheme to detect the error.

\begin{figure*}
        \centering
       \includegraphics{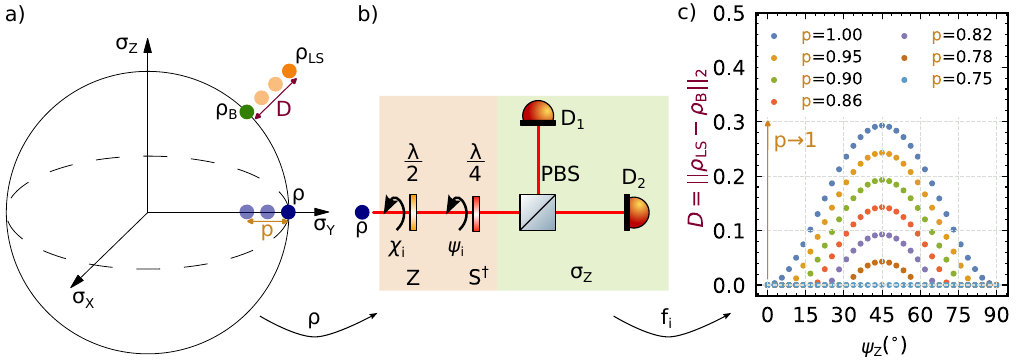}
    \caption{Illustration of the detection scheme for systematic errors 
    in a simple setting. a) The Bloch sphere is the physical state space 
    of a single qubit. The blue dot denotes a quantum state $\rho$ to be determined from  Pauli measurements. 
    Severe systematic or statistical error can lead to a least-squares estimate $\rhoLS$ (orange dot) outside the Bloch sphere, and thus the distance $D$ 
    (purple arrow) between a physical (but biased) estimate 
    $\rhoBiased$ of Eq.~(\ref{def:PhysicalEstimator}) (green dot) 
    is finite. If this distance $D$ persists for large statistical samples, 
    this is a signature of a systematic error. The distance $D$ decreases with lower purity $p=\trace(\rho^2)$ of the state $\rho$, as suggested by the purple and yellow arrows.  b) Single-qubit Pauli measurements for polarized photons: in the brown area the $k$-th Pauli measurement setting for obtaining the two frequencies $f_i$ and $f_j$, corresponding to the eigenvalues, is adjusted by the rotations $\chi_{k}$ and $\psi_{k}$ of the half-wave plate ($Z$ gate) and the quarter-wave plate ($S^{\dagger}$ gate), respectively. The polarizing beam splitter (PBS) and single-photon detectors (D1 and D2) conduct the $\PauliZ$ measurement, shown in the green area. c) The calculated distance $D$ depends on the adjustment angle $\psi_{Z}$ of the Pauli $Z$ basis, and its magnitude decreases with the purity of the true, underlying state $\rho=\ket{\mathrm{i}}\bra{\mathrm{i}}$ ($\PauliY$ eigenstate). The maximum $D$ at $\psi_{\text{Z}}=\SI{45}{\degree}$ occurs when $\PauliY$ is measured instead of $\PauliZ$, and for a purity $p=0.75$ the error cannot be detected anymore.}
    \label{fig:SingleQubit}
\end{figure*}

\section{Systematic error detection method}
We first review measurement schemes for tomography and quantum state estimators, as they form the basis for the systematic error detection method introduced subsequently.
\subsection{Measurement schemes for tomography}
To determine the quantum state, one measures a set of tomographically complete measurement operators $\operatorname{P}_i$ on $N$ copies of the quantum state, described by the density operator $\rho$. The operators $\operatorname{P}_i$ are Hermitian and constitute a basis to the space of all observables where $i$ denotes a specific measurement outcome. Local Pauli measurements~\cite{KwiatPRA2001} are a frequently used tomographically complete set of measurements for qubits, consisting of all $3^n$ strings of Pauli measurements acting on $n$ qubits. From now on, we focus on these, albeit our idea works for other measurement schemes~\cite{KaszlikowskiPRA2004, Teo_2010, ZhuPRA2011, StrickerPRX2022, jung2024coarsegrained} as well. Quantum state tomography determines $\rho$ from the experimentally observed frequencies $f_i$, which approximate the outcome probabilities and are defined as the number of times the outcome of a tomographically complete set of observables $\operatorname{P}_i$ occurs relative to $N$. In the limit of $N\to \infty$ one recovers the true underlying outcome probabilities as predicted by Born’s rule.

\subsection{Biased and unbiased state estimators}
Once the outcome frequencies are obtained from the measurement procedure, an estimate for the quantum state can be determined. For that, different estimator kinds exist, which may be unbiased or biased. By definition, unbiased estimators have the property that the expectation value of the estimator equals the true value. From the fact that the set of allowed density matrices is constrained by the positivity of their eigenvalues, one can show that unbiased estimators deliver 
necessarily nonphysical states~\cite{SchwemmerPRL2015}, meaning that the predicted state $\rhoUnbiased$ may be not positive semidefinite.

The least-squares estimator is an unbiased estimator, which we focus on and is defined for single qubits as:
\begin{align}
\rhoLS = 
\frac{1}{2}\Big(\openone+ \sum^{3 \cdot 2 }_{i=1} f_i P_i \Big), \label{eq:defLeastSquares}
\end{align}
where $f_i$ are the frequencies observed experimentally. For $n$-qubit states, Eq.~(\ref{eq:defLeastSquares}) sums over the $4^n-1$ non trivial tensor products of Pauli measurements, each with two possible outcome frequencies. Frequency $f_i$ can occur where the resulting estimate $\rhoLS$ has negative eigenvalues. This is more likely if the underlying quantum state $\rho$ has a high purity and lies close to the surface of the Bloch sphere, see Fig.~\ref{fig:SingleQubit}\textcolor{blue}{a}.

On the other hand, to ensure valid quantum states, one may impose additional positivity constraints. This necessarily leads to biased state estimators. For example, one may take as
an estimate $\rhoBiased$ the closest positive semidefinite state to the least-squares estimate $\rhoLS$ with respect to the squared Hilbert-Schmidt norm~\cite{Guta_2020}, which can be expressed as the optimization problem,
\begin{align}
        \rhoBiased = \argmin _{\substack{\operatorname{Tr}(\varphi) = 1 \\ \varphi \succcurlyeq 0}} ||\rhoLS-\varphi||^{2}_2. \label{def:PhysicalEstimator}
\end{align} 
Clearly, if only statistical fluctuations are present, the unbiased estimator $\rhoLS$ converges to a physical state in the limit of many repetitions of the experiment $N\rightarrow \infty.$ If, however, a severe systematic error occurs, this is not necessarily the case. One may end up with very different estimates $\rhoLS$ and $\rhoBiased$, which results in a non-negligible distance $D$ between the estimates,
\begin{align}
    D = ||\rhoLS-\rhoBiased||_2=\operatorname{Tr}\Big[(\rhoLS-\rhoBiased)^2\Big]^{1/2}, \label{def:Distance}
\end{align}
which utilizes the Hilbert-Schmidt norm as a metric. If $D > 0$ even for large $N$, this is a signature of systematic errors. This is the core idea of our method to recognize systematic errors.

Fig.~\ref{fig:SingleQubit} illustrates the concept of our systematic error detection method for single-qubits. In Fig.~\ref{fig:SingleQubit}\textcolor{blue}{a}, we see the Bloch sphere, the physical state space of a single qubit, and as blue dots the state $\rho$. We assume that a severe systematic or statistical error occurs (for demonstration purposes), which leads to a least-squares estimate $\rhoLS$ (orange dot) far outside of the Bloch sphere. Consequently, the distance $D$, purple arrow, between the biased estimate $\rhoBiased$, green dot, is non-negligible. If we apply a depolarization channel to $\rho$, its purity $p$ reduces, indicated by the yellow arrow between the blue dot and the transparent blue dots. With decreasing purity $p$, the distance $D$ also decreases (see transparent orange dots).  

Fig.~\ref{fig:SingleQubit}\textcolor{blue}{b} shows the single-qubit Pauli measurement setting for qubits implemented by the polarization of photons. The $k$-th basis measurement setting determines the frequencies $f_i$ and $f_j$ that correspond to its eigenvalues, and is adjusted in the brown area by the rotations $\chi_{k}$ and $\psi_{k}$ of the phase gates $Z$ and $S^{\dagger}$. The phase gates $Z$ and $S^{\dagger}$ correspond to a half- ($\lambda /2$) and quarter-wave ($\lambda /4$) plate, respectively. Subsequently, the measurement $\PauliZ$ is performed in the green area, which is implemented by a polarizing-beam splitter (PBS) followed by two photon detectors (D1, D2). Fig.~\ref{fig:SingleQubit}\textcolor{blue}{c} shows the calculated distance $D$ from Eq.~(\ref{def:Distance}) as a function of the basis adjustment angle $\psi_{\text{Z}}$ for the state $\rho=\ket{\mathrm{i}}\bra{\mathrm{i}}$, $+1$ eigenstate of $\mathrm{Y}$, where an applied depolarizing channel tunes the purity from $0.75$ to $1$. We simulate all Pauli measurements as observed correctly, except for $\PauliZ$, where we vary the alignment angle $\psi_{\text{Z}}$, with $\SI{0}{\degree}$ indicating perfect alignment. For $\psi_{\text{Z}} = \SI{45}{\degree}$, we find the maximum distance $D$ when measuring $\PauliY$ instead. The distance $D$ decreases with state purity and reaches zero at a purity of $p_{\text{min}}=0.75$, see Appendix~\ref{app:criticalPurity} for details.

\subsection{Rigorous formulation of the method}
The definition of $D$ alone cannot certify systematic errors, as it does not differentiate between statistical and systematic errors. Concentration bounds have proven to be useful statistical tools to find the probability that a quantity obtained by finite experimental repetitions deviates from its true mean. We follow the spirit of Ref.~\cite{degois2023userfriendly}, and interpret quantum state tomography as sum of independent, zero-mean, random vectors such that we can utilize the vector Bernstein inequality to derive the following proposition, see Appendix~\ref{lab:statisticalMeaning} for details. 
\paragraph*{Proposition 1.}
Let $\rhoLS$ and $\rhoBiased$ be the least-square and biased estimate from Eq.~(\ref{eq:defLeastSquares}) and Eq.~(\ref{def:PhysicalEstimator}), respectively. The observed frequencies $f_i$ determine the least-squares estimate $\rhoLS$ from measuring {$N$} identical state copies of the underlying quantum state $\rho$. Then, the probability $\delta_{\text{sta}}$ that the distance $D$ of Eq.~(\ref{def:Distance}) between $\rhoLS$ and $\rhoBiased$ is equal to or greater than $\tau$ obeys:
\begin{align}
 \delta_{\text{sta}}=\mathbb{P}\left[D \geqslant \tau\right] \leq 8 \exp \left[-\frac{{N} \, \tau^2}{2 \times 5^n} \frac{3}{3+\sqrt{2} \tau / \sqrt{5^n} }\right], \label{eq:systematicProbability}
\end{align}
where $n$ is the number of qubits on which local Pauli measurements are measured.

This bounds the probability $\delta_{\text{sta}}$ that only statistical 
errors cause a distance $D>\tau$ between $\rhoLS$ and $\rhoBiased$,
and if this is smaller than a threshold fixed beforehand, one can reject
the hypothesis that only statistical errors occur.
Consequently, if the distance $D$ of Eq.~(\ref{def:Distance}) is significantly larger than the corresponding $\tau$, the systematic error test successfully detects a systematic error with confidence level $1-\delta_{\text{sta}}$.

\begin{figure*}
        \centering
       \includegraphics{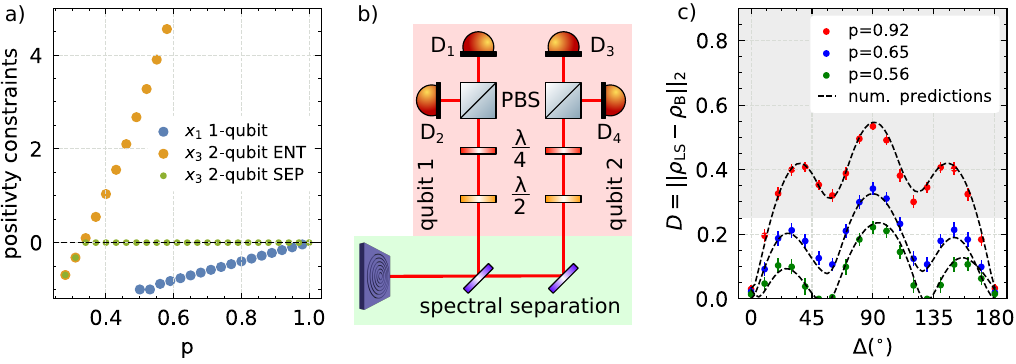}
    \caption{a) Capability of single-qubit states (blue) and two-qubit states (entangled in orange, separable in green) to detect the error of exchanging $\PauliY$ and $\PauliZ$ on qubit one. We show the maximum positivity constraint values depending on the purity $p$ of the underlying state $\rho$, and find that only entangled states yield values $x_3$ above the dashed line, meaning that they detect this error. b) The two entangled photons from the quantum dot are prepared and separated in the green area, and individual single-qubit Pauli measurements are performed in the red area, see Refs.~\cite{juliaNeuwirthPRB,rota2024source} for details. c) We show our experimental results for the distance $D$ depending on $\Delta$, the offset on qubit one that we apply for each measurement setting $k$, i.e. $\psi_{k} = \psi_{k}^{\mathrm{theo}}+\Delta$. The red, blue and green dots are in accordance with the lines from the simulated predictions for all three quantum states, and correspond to purities from $0.56$ to $0.92$. The confidence level to detect a systematic error begins with $90\%$ for a distance $D \geq 0.25$ (grey area), which allows detection of almost all errors for highly entangled states (red points). }
    \label{figure2}
\end{figure*}

\section{Necessary state properties for error detection}
The single-qubit example above highlights that the purity of the state is relevant to detect errors. So, we investigate the minimal purity, for which there exists a one- or two-qubit state $\rho$, which detects a local single-qubit error on qubit one. We assume that the systematic error on qubit one is such that the corrupted Pauli measurements $\tilde{\sigma}_\mu$ are linear combinations of all three Pauli measurements:
\begin{align}
    \tilde{\sigma}_{\mu} = \sum^{3}_{\nu=1} \operatorname{M}_{\mu \nu} \sigma_{\nu}, \label{eq:errorSigmaMatricesMain}
\end{align}
where the rows of the misalignment matrix $\operatorname{M}$ have to be normalized to ensure that the expectation values of the Pauli observables are restricted to $\pm 1$. The error $\operatorname{M}$ influences all observed frequencies on qubit one and, as a consequence from Eq.~(\ref{eq:defLeastSquares}), we obtain a corrupted least-squares 
estimate $\tilde{\rho}_{\text{LS}}$. 

To check whether the least-squares estimate corresponds to a valid state, we use results on the positivity constraints of multi-qubit states~\cite{KIMURA2003339, GamelPRA}. Any quantum state $\rho$ has positive eigenvalues and must satisfy the inequalities $\zeta_l(\rho) \leq 0$~\cite{KIMURA2003339, GamelPRA}. The number of inequalities, $l=2^n-1$, depends on the number of qubits $n$. If any of the parameters $\zeta_l(\tilde{\rho}_{\text{LS}})$ is greater than zero, then $\tilde{\rho}_{\text{LS}}$ has a negative eigenvalue, and one can use our method to detect the error. We begin with the minimal necessary purity for single-qubit systems in the following.

\subsection{Single-qubit system}\label{sect:singleQubitPurity}
For the single-qubit case, we analytically determine the minimal necessary purity $p_{\text{min}}$. The actual underlying state is $\rho_{\text{LS}}=u_i \sigma_i/2$, which has a purity of $\operatorname{Tr}(\rho^2) = u^2_i/2$ with $i \in \{0,3\}$ and $u_0=1$.  If we suppose to have the misalignment $\operatorname{M}$, the erroneous least-squares estimate results in $\tilde{\rho}_{\text{LS}} =\tilde{u}_i \sigma_i/2$ with $i \in \{0,3\}$. As a remark, we assume that the coefficient in front of $\sigma_0=\openone$, is not affected by errors; hence $u_0=\tilde{u}_0=1$. For single-qubit states only one positivity constraint $\zeta_1$ exists:
\begin{equation}
    \zeta_1(\rho) = \trace{\rho^2} -1 \leq 0 \label{eq:posConstrZeta}.
\end{equation}
To determine the minimal necessary purity of $\rho$, we begin with the minimal purity $\wp = \trace(\rho^2) = 1/2$ of the maximally mixed state $\rho = \openone/2$ and determine $\tilde{\rho}$ that maximizes Eq.~(\ref{eq:posConstrZeta}). We then increase the variable purity $\wp$ of the underlying state until we find a $\tilde{\rho}$ that violates Eq.~(\ref{eq:posConstrZeta}), which we express as an optimization problem depending on $\tilde{\rho}$ and $\operatorname{Tr}(\rho^2)$:
\begin{align}
x_1 =& \max_{\substack{ \operatorname{Tr}(\rho^2) \leq \wp} } \zeta_1(\Tilde{\rho}_{\text{LS}}(\operatorname{M}))=\max_{\substack{ u_i \sigma_i/2 \leq \wp} } (\tilde{u}^2_i)/2 -1 \nonumber\\
=&\max_{\substack{ \sum_{i}u^2_i\leq \Bar{u}}} ||  \operatorname{M}^T \bold{u}||^2 =~\Bar{u} \, \sigma^2_{\text{max}}(\operatorname{M}^T),
\label{eq:singleQubitMaximization}
\end{align}
where $\sigma_{\text{max}}$ denotes the largest singular value of the misalignment matrix $\operatorname{M}$, and $\Bar{u}$ is the absolute value of the true underlying Bloch vector $\bold{u}$.

We define the minimal necessary purity $p_{\text{min}}$ as the first purity $\wp$ that violates the positivity condition given by Eq.~(\ref{eq:posConstrZeta}), and in this case the maximal absolute value of the Bloch vector $u_{\text{max}}$ is equal to one, \mbox{$1=u_{\text{max}}\, \sigma^2_{\text{max}}(\operatorname{M}^T)$}. From this we determine the lowest necessary purity $p_{\text{min}}$ as follows:
\begin{align}
    p_{\text{min}} &= \frac{1}{2}\left(1+u_{\text{max}}\right)=  \frac{1}{2}\left(1+\frac{1}{\sigma^2_{\text{max}}(\operatorname{M}^T)}\right). \label{eq:singlePurityFormula}
\end{align}
If we consider the numerical results in Fig.~\ref{fig:SingleQubit}\textcolor{blue}{c}, the maximum distance $D$ is obtained for $\psi_{\text{Z}}=\SI{45}{\degree}$, where $\PauliY$ is measured instead of $\PauliZ$, which corresponds to the following misalignment matrix $\operatorname{M_Z}$:
\begin{equation}
    \operatorname{M_Z}=\left(\begin{array}{lll}
1 & 0 & 0 \\
0 & 1 & 0 \\
0 & 1 & 0
\end{array}\right).
\label{endmatter:MZ}
\end{equation}
We determine the minimal necessary purity $p_{\text{min}}=0.75$ from Eq.~(\ref{eq:singlePurityFormula}) by plugging in $\operatorname{M_Z}$, which is in perfect agreement with our numerical results in Fig.~\ref{fig:SingleQubit}\textcolor{blue}{c}. Now, we extend our discussion to two-qubit systems.

\subsubsection{Two-qubit system}
We assume that the errors are local systematic errors, which means that there is no correlated error. Thus, the first and second qubits may be subject to local errors $\operatorname{M}_1$ and $\operatorname{M}_2$, respectively. We express the least-squares estimate in the Einstein-sum convention by $\rhoLS =1/4~r_{\mu \nu}~ \sigma_{\mu} \otimes \sigma_{\nu}$, where $\mu, \nu \in \{0,3\}$, where we assume that the coefficient $r_{00} = 1$. We change our notation, as we directly use the results of Ref.~\cite{GamelPRA}, which collects these coefficients $r_{\mu \nu}$ in the Bloch matrix $\stackrel{\leftrightarrow}{r}$: 
\begin{equation}
\stackrel{\leftrightarrow}{r}=\left[\begin{array}{c|lll}
1 & r_{01} & r_{02} & r_{03} \\
\hline r_{10} & r_{11} & r_{12} & r_{13} \\
r_{20} & r_{21} & r_{22} & r_{23} \\
r_{30} & r_{31} & r_{32} & r_{33}
\end{array}\right] \equiv\left[\begin{array}{cc}
1 & \bold{v}^{\dagger} \\
\bold{u} & \operatorname{R}
\end{array}\right],
\end{equation}
where the vectors $\bold{u}$ and $\bold{v}$ correspond to the local Bloch vectors of the first and second qubit, respectively, and $\operatorname{R}$ is the correlation matrix. From the misalignment matrices we obtain the erroneous local Bloch vectors \mbox{$\Tilde{\bold{u}}=\operatorname{M}_{1}\bold{u}$} and \mbox{$\Tilde{\bold{v}}=\operatorname{M}_{2}\bold{v}$}, and the erroneous correlation matrix \mbox{$\Tilde{\operatorname{R}}=\operatorname{M}_{1}\operatorname{R}\operatorname{\,M}^T_{2}$}; see details in the Appendix~\ref{app:criticalPurity}. 

In the two-qubit case we have three positivity constraints $\zeta_i$, which depend on the local Bloch vectors and the correlation matrix as follows~\cite{GamelPRA}:
\begin{align}
 &\zeta_1(\rho)= \|\stackrel{\leftrightarrow}{r}\|^2 -4 \leq 0 \\
 &\zeta_2(\rho)=  \left(\|\stackrel{\leftrightarrow}{r}\|^2-2\right) - 2\left(\bold{u}^{\dagger} \operatorname{R} \bold{v}-\operatorname{det} \operatorname{R}\right) \leq 0 \\
 &\zeta_3(\rho)=\nonumber \\ \nonumber &-8\left(\bold{u}^{\dagger} \operatorname{R} \bold{v}-\operatorname{det} \operatorname{R}\right)-\left(\|\stackrel{\leftrightarrow}{r}\|^2-2\right)^2-8 \bold{u}^{\dagger} \operatorname{R}_{\text{cof}} \bold{v} \nonumber \\ &+4\left(\|\bold{u}\|^2 \, \|\bold{v}\|^2+\left\|\bold{u}^{\dagger} \operatorname{R}\right\|^2+\|\operatorname{R} \bold{v}\|^2+\|\operatorname{R}_{\text{cof}}\|^2\right) \leq 0, \label{app:eq:2qubitzetacondition}
\end{align}
where $\operatorname{R}_{\text{cof}}$ is the cofactor matrix of $\operatorname{R}$. The entries of the cofactor matrix $\operatorname{R}_{\text{cof,\,}i,j}$ are obtained from the first minor (determinant for square matrices) of the matrix that is obtained from $\operatorname{R}$ by deleting $i$-th row and $j$-th column, which are, in addition,  multiplied by $(-1)^{i+j}$. The term ${\|\stackrel{\leftrightarrow}{r}\|^2} = 1+\|\bold{u}\|^2+\|\bold{v}\|^2+\|\operatorname{R}\|^2$ is proportional to the purity of the state, and the norms of the Bloch vector and the correlation matrix are the Euclidean and Hilbert-Schmidt inner product $\|\operatorname{R}\|^2=\operatorname{Tr}\left(\operatorname{R}^{\dagger} \operatorname{R}\right)$, respectively. 

The idea of determining the minimal necessary state purity $p_{\text{min}}$ of $\rho$ for single-qubit states of Appendix~\ref{sect:singleQubitPurity} naturally extends to two-qubit states. We assume that both qubits are subject to the misalignments $\operatorname{M}_1$ and $\operatorname{M}_2$, respectively. To determine the minimal necessary purity of the underlying state $\rho$, we increase the purity $\wp$ of $\rho$, beginning with the maximally mixed two-qubit state. We then ask whether for the estimate $\tilde{\rho}_{\text{LS}}$ one of the positivity constraints is violated. We denote by $x_l$ the maximum violation that can be reached by any constraint $\zeta_l(\tilde{\rho}_{\text{LS}})$:
\begin{align}
    x_l=\max_{\substack{ \rho \;{\rm with}\;\trace{(\rho^2)}\leq \wp, \\ \{\zeta_i(\rho) \leq 0\} \;{\rm for }\; i=1,2,3}} 
    \zeta_l(\tilde{\rho}_{\text{LS}}) \quad \forall l \in \{1,2,3\}.
    \label{eq:twoQubitMaximization}
\end{align}
Note that we use all three positivity conditions $\zeta_i$ to ensure that the true underlying two-qubit state $\rho$ is positive semidefinite. We begin with a state purity of $\wp=0.25$ and increase it until one of the results $x_l$ becomes positive. The corresponding value of $\wp$ is then defined as the minimum required purity of the state, $p_{\text{min}}$.
\begin{figure}
    \centering
    \includegraphics{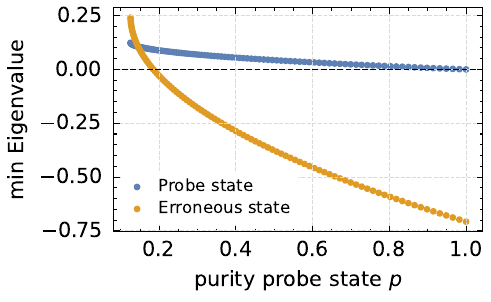}
    \caption{We consider an error of measuring $\PauliY$ instead of $\PauliZ$ on the first qubit of the three-qubit state probe state $\psi_{\text{probe}}$ at which we apply white noise Eq.~(\ref{eq:whitenoise}) to change the purity $p$. The orange and blue dots represent the smallest eigenvalue of the erroneous and probe states, respectively, as a function of $p$. The minimal necessary purity for our error detection method is $p^{\text{appr}}_{\text{min}}=0.18$, where the erroneous estimate has a negative eigenvalue as marked by the dashed black line. In comparison, the probe state does not have any negative eigenvalues.}
    \label{fig:minEig_ProbeState}
\end{figure}

Inspired by this analysis, we further investigate the necessary state purity to detect errors. In particular, we want to see whether the detection capabilities of product and entangled states differ. To compare the error detection capability of entangled and product states, we add to Eq.~(\ref{eq:twoQubitMaximization}) additional conditions $\varepsilon_i$ from the PPT criterion~\cite{PeresPRL_1996_PPT,HORODECKI_PLA_1996}. The PPT criterion for qubit systems states that if $\rho$ is separable, then its partial transpose on the first (second) subsystem $\rho^{\operatorname{T}_{1(2)}} \geq 0$ must be positive semidefinite. If we insert $\rho^{\operatorname{T}_2}$ into Eq.~(\ref{app:eq:2qubitzetacondition}), we get conditions $\varepsilon_i$ that ensure the positivity of the partial transpose of $\rho^{\operatorname{T}_2}$. Since $\PauliY$ is the only Pauli matrix that changes sign under partial transposition, only the terms $\operatorname{det} \operatorname{R}$ and $8 \bold{u}^{\dagger} \operatorname{R}_{\text{cof}} \bold{v}$ in Eq.~(\ref{app:eq:2qubitzetacondition}) change signs and result in two additional conditions $\epsilon_2$ and $\epsilon_3$; see Appendix~\ref{app:criticalPurity} for details. If we restrict ourselves to separable states, the optimization problem Eq.~(\ref{eq:twoQubitMaximization}) changes to:
\begin{align}
    x_l=\max_{\substack{ \trace{(\rho^2)}\leq \wp, \\ \{\zeta_1(\rho) \leq 0,  \zeta_2(\rho) \leq 0, \zeta_3(\rho) \leq 0 \} \\ \{ \varepsilon_2(\rho) \leq 0, \,\varepsilon_3(\rho) \leq 0 \}} } \zeta_l(\tilde{\rho}_{\text{LS}}) \quad \forall l \in \{1,2,3\}.
    \label{eq:twoQubitSEPMaximization}
\end{align}

Concretely, we analytically determine the minimal necessary state purity $p_{\text{min}}$ for all possible misalignment errors $\operatorname{M}_1$ in the $\PauliY$ and $\PauliZ$ basis acting on qubit one:
\begin{align}
    \operatorname{M}_1=\left(\begin{array}{lll}
    1 & 0 & 0 \\
    \sin{(\beta)}\sin{(\alpha)} & \cos{(\alpha)} & \cos{(\beta)}\sin{(\alpha)} \\
    \sin{(\delta)}\sin{(\gamma)} & \cos{(\delta)}\sin{(\gamma)} & \cos{(\gamma)}
\end{array}\right),
\label{eq:errorSettingSearchMain}
\end{align} 
depending on the four angles $\alpha$, $\beta$, $\gamma$, and $\delta$. Note that this is indeed already the most general error on a single qubit in the formalism of misalignment matrices, since rotations of the $\PauliX$ measurement can be counteracted by a local unitary transformation.

We determine $p_{\text{min}}$ of single-qubit states, separable two-qubit states, and general two-qubit states for all combinations $\alpha, \beta, \gamma, \delta$ of $\{0,\frac{\pi}{8},\frac{\pi}{4},\frac{3\pi}{8},\frac{\pi}{2}\}$. First, we find, without any exception, that for two-qubit quantum states, less purity is sufficient for the detection of systematic errors compared to single-qubit systems. The values of $p_{\text{min}}$ for two-qubit systems are between $0.29$ and $0.34$ which is significantly below the smallest possible purity of the maximally mixed single-qubit state $0.5$. Still, this result has to be handled
with care, as purities of states with a different number of qubits are difficult to compare.

Furthermore, we find that the error of exchanging $\PauliY$ and $\PauliZ$, corresponding to a partial transposition on qubit one, gives different results for entangled and product states. In particular, only entangled states can detect this error. In Fig.~\ref{figure2}\textcolor{blue}{a} we show the results for the maximum positivity constraints $x_l$ depending on the state purity $p$ where the error is the described exchange on the first qubit. The orange and green dots are the $x_3$ values for entangled and separable states, respectively, and the values $x_1$ for single qubits are represented by the blue dots. Only the values of the entangled state exceed the black dashed line at $x_l=0$ at $p_{\text{min}}=0.33$, which means that there exists a state that is suitable for detecting the error. We stress that an unintentional exchange of the wave plates in Fig.~\ref{fig:SingleQubit}\textcolor{blue}{b} is an example of a partial transposition error and thus 
it can only be detected by entangled states. 

In principle, it is possible to determine $p_{\text{min}}$ for $n$-qubit states, but the number of optimization variables scales as $2^n \times 2^n$. Therefore, in the following, we present an approximation for determining the $n$-qubit critical purity.

\subsubsection{Approximation for $n$-qubit critical purity}\label{app:criticalPurity}
Here, we propose an approximation that utilizes the fact that many quantum states observed in the laboratory can be accurately described as pure states affected by random noise. Thus, we propose to search for a pure probe state $\rho_{\text{probe}}=\ket{\psi}\bra{\psi}$ with components $\psi_i = a_i + \mathrm{i}\, b_i$
instead of an arbitrary state $\rho_{\text{probe}}$. Now, the optimization problem depends on only $2\times 2^{n}$ variables. We select the probe state $\rho_{\text{probe}}$ as the state for which the erroneous least-squares estimate $\Tilde{\rho}_{\textrm{LS}}$ exhibits the largest negative eigenvalue:
\begin{align}
 \max_{\substack{ \sum_{i} a^2_i+ b^2_i = 1} } -\operatorname{spec}(\Tilde{\rho}_{\textrm{LS}}),
\label{eq:optimizationProbe}
\end{align}
where we demand normalization of the probe state $\ket{\psi}$. We apply white noise with strength $\epsilon_{\text{dep}}$ to the probe state:
\begin{align}
    \rho^{\text{noise}}_{\text{probe}} = (1-\epsilon_{\text{dep}})\rho_{\text{probe}} + \epsilon_{\text{dep}} \frac{\openone}{2^n}
    \label{eq:whitenoise}
\end{align}
until we find any negative eigenvalue. We define the minimal necessary purity $p_{\text{min}}$ as the first purity of $\tilde{\rho}_{\text{probe}}$, such that all eigenvalues of $\tilde{\rho}_{\text{probe}}$ are positive.

We test the approximation for the case of measuring $\PauliY$ instead of $\PauliZ$ as shown in Fig.~\ref{fig:SingleQubit}\textcolor{blue}{c} and Fig.~\ref{figure2}\textcolor{blue}{a}. For single-qubit states we get with this approximation a critical purity of $p^{\textrm{appr}}_{\text{min}}=p_{\text{min}}=0.75$, which is the same as we find in Fig.~\ref{fig:SingleQubit}\textcolor{blue}{c} and from Eq.~(\ref{eq:singlePurityFormula}). We find for a two-qubit state with this error on the first qubit a purity of $p^{\text{appr}}_{\text{min}}=0.38$. The two-qubit purity is slightly higher than the exact value $p^{\text{appr}}_{\text{min}}=0.33$, but is still much smaller than the purity of states commonly used for quantum information tasks. 

For three qubits having the mentioned error at qubit one, we find a minimal necessary purity of $p^{\text{appr}}_{\text{min}}=0.18$ with this approximation. Fig.~\ref{fig:minEig_ProbeState} shows the minimum eigenvalue as a function of the purity $p_{\text{probe}}$. The dashed black line highlights the zero eigenvalue line, and if an estimate has a negative eigenvalue, we can detect it with our method of Eq.~(\ref{def:Distance}).

\section{Experimental demonstration}
We demonstrate that our method is practically feasible by using entangled two-qubit photonic states emitted by a GaAs quantum dot~\cite{da2021gaas}. Our device~\cite{rota2024source}, which embeds the quantum dot, achieves high coincidence rates with an integrated cavity and allows tuning the purity of the quantum state by strain tuning of the excitonic fine structure of the emitter~\cite{da2021gaas,rota2024source,trottaPRl2015}. The achieved coincidence rate allows one to perform quantum tomography as a function of a measurement variable and collect enough data to properly estimate the statistical uncertainty while ensuring stability of the experimental conditions during acquisition. 
Fig.~\ref{figure2}\textcolor{blue}{b} illustrates the experimental setup, which includes the light emission from the quantum dot followed by spectral filters to split the state into two single-qubit measurement paths. We consider the scenario where all measurement settings $k$ on qubit one have a constant angular offset $\Delta$ in the quarter-wave plate adjustment angle $\psi_{k} = \psi_{k}^{\mathrm{theo}}+\Delta$. This systematic error is relevant in laboratory practice because it corresponds to an error in the alignment of the wave plates.

Fig.~\ref{figure2}\textcolor{blue}{c} shows the experimental results for distance $D$ depending on the offset $\Delta$ for quantum states with three different purity values from $0.56$ to $0.92$ that result from a single strain-tuned quantum dot~\cite{trottaPRl2015,da2021gaas,rota2024source}. The distance $D$ increases as a function of the purity, which agrees with the previous discussion. The black line is the simulation of the expected distance $D$ as a function of $\Delta$ by using the correctly measured quantum states at $\Delta=0$. We see that the simulation agrees with the measurement result. We underline that our method detects systematic errors with a confidence level of $90\%$ or higher for distances $D\geq 0.25$, which is reached for nearly all errors $\Delta$ for highly entangled states. We refer to Appendix~\ref{app:dataEvaluation} for details on the statistical evaluation and the confidence level of the experimental results. 

\section{Discussion}
Systematic errors in the readout of quantum information may be severe and complex to recognize. We have introduced a user-friendly error detection method that compares the estimates from an unbiased and biased estimator on the same measurement data. Additionally, we have determined the probability that any deviation between these estimates is due to a systematic error by taking into account the number of measurement repetitions. Furthermore, we have analytically determined the minimum state purity required to detect a single-qubit error and found that entangled states are particularly effective for this purpose. Finally, we have experimentally demonstrated the applicability of our error detection method for Pauli measurements of a two-qubit photonic state. Our results show that this method is valuable for identifying systematic measurement errors and that the purity of the quantum state affects the error detection capabilities. We provide a non-exhaustive discussion on the detection of other common systematic errors on qubits implemented by the photon's polarization in Appendix~\ref{app:furtherErrors}.

An experimentally relevant extension of this work is the analysis of the role of correlated errors. In this scenario, it is especially interesting to study the performance of entangled states. Future studies on the role of multiple single-qubit errors are also highly desirable. In fact, one can show that two errors may cancel each other such that not even an entangled state would be sufficient to detect it, see Appendix~\ref{app:criticalPurity} for examples. Closely related to this aspect is the search for a suitable modification of our method that enables the location of the qubit on which the error occurs. We predict that this work and its future developments will help researchers design their experiments and offer a valuable tool for the faithful implementation of quantum information processing and quantum communication protocols.

\section*{Acknowledgments}
JF is grateful to W. Dür for supporting this project and thanks T. Kraft, C. de Gois, T. Fromherz, C. Schimpf, and J. Neuwirth for fruitful discussions. JF's research is funded in whole or in part by the Austrian Science Fund (FWF) 10.55776/P36010. For open access purposes, the author has applied a CC BY public copyright license to any author accepted manuscript version arising from this submission. OG was 
supported by the Deutsche Forschungsgemeinschaft  (DFG, German Research Foundation, project numbers 447948357, 440958198 and 63437167), the Sino-German Center for Research Promotion (Project M-0294), and the German Ministry of Education and Research (Project QuKuK, BMBF Grant No.\ 16KIS1618K and Project BeRyQC, Grant No. 13N17292)). The project is funded within the QuantERA II Programme that has received funding from the EU H2020 research and innovation programme under GA No. 101017733 via the project QD-E-QKD. The authors also acknowledge support from MUR (Ministero dell’Università e della Ricerca) through the PNRR MUR project PE0000023-NQSTI and the European Union’s Horizon Europe research and innovation program under EPIQUE Project GA No. 101135288, Ascent+ project GA No. 871130, the FWF via the Research Group FG5 [10.55776/FG5], and the cluster of excellence quantA [10.55776/COE1] as well as the Linz Institute of Technology (LIT) and the LIT Secure and Correct Systems Lab, supported by the State of Upper Austria. THL acknowledges funding from the BMBF through the Project Qecs (FKZ: 13N16272). S.F.C.D.S. acknowledges funding from the São Paulo Research Foundation (FAPESP Process Numbers 2024/08527-2 and 2024/21615-8).

\section*{Author contributions}
JF and OG developed the theory. JF implemented the source code and performed the calculations. JF, FBB, TMK, MBR, QB, SFCdS, SS, SH, THL, AR and RT worked at the design and fabrication of the QD sample. The measurements were conducted by FBB, AL, and MB, and RT coordinated the measurements. JF analyzed the results with the help of FBB. JF and OG coordinated the project. RK provided conceptual guidance/assistance throughout the early stages of this project. JF and OG wrote the manuscript with the help of FBB, TMK, and RT. All authors provided fruitful suggestions for improving the manuscript.

\section*{Availability of data and source code}
The measurement data and source code that support the results of this work are provided upon reasonable request to FBB and JF, respectively.

\clearpage
\appendix
\section{Quantum State Estimators\label{lab:reviewEstimators}}
In this section, we briefly review the definition of a tomographically complete set of measurements. We continue with a summary on biased and unbiased quantum state estimators. In particular, we focus on the least-squares estimator and the convex optimization problem returning the closest physical estimate to the linear inversion estimate, as we have introduced in the main text. To provide the reader with alternatives for state estimators, we briefly explain the maximum-likelihood estimators.

\subsection{Tomographically complete set of measurements}\label{app:tomoCompleteSet}
A tomographically complete set of measurements is a set of Hermitian operators that constitutes a basis to the operator space $\mathcal{L}(\mathbb{H}_d)$, and allows to fully reconstruct the pre-measurement state from its expectation values. For instance the set of Pauli observables $\{\sigma_{\operatorname{X}},\sigma_{\operatorname{Y}},\sigma_{\operatorname{Z}}\}$ is an example for a tomographically complete set of measurements on a single qubit. One can write a tomographically complete set of operators as an informationally complete positive operator-valued measure (IC POVM), which is a set of Hermitian, positive semidefinite operators $\operatorname{P}_i \in \mathbb{H}_d$, which return unique probability distributions for two different states $\rho\neq \sigma$. In order to normalize the returned probability distributions we demand that $\sum^{m}_{i=1} \operatorname{P}_i = \openone$ holds. The measurement of a quantum state $\rho \in \mathcal{L}(\mathbb{H}_d)$ results in one of the outcomes $i \in \left\{1, \dots m\right\}$ and the probability $[p]_i$ that outcome $i$ is observed is given by Born's rule:
\begin{equation}
    [p]_i = \trace (\operatorname{P}_i \rho), \label{Bornsrule}
\end{equation}
where $[p]_i$ denotes the $i$-th component of the probability vector $p \in \mathbb{R}^m$. It is impossible to directly access the probabilities, Eq.~(\ref{Bornsrule}), from a single copy of $\rho$, but by performing a sequence of measurements on $N$ identically prepared copies of $\rho$ one can estimate outcome frequencies $f_N$ for the probabilities given by Born's rule:
\begin{equation}
    [f_N]_i = \frac{n_i}{N}, \label{frequencies}
\end{equation}
where $n_i$ is the number of times the outcome $i$ is observed and in the limit of $N \rightarrow \infty$ the observed frequencies correspond to the true probabilities. 

\subsubsection{Single-qubit Pauli basis measurements}
A widely used tomographically complete set of measurements are local single-qubit Pauli basis measurements. The set of operators $\operatorname{P}_i$ for Pauli basis measurements written as an IC POVM on a single-qubit system is as follows:
\begin{align}
   \operatorname{P}_i= \left\{\frac{1}{3}\ket{\sigma^{+}_{\mathrm{X}}}\bra{\sigma^{+}_{\mathrm{X}}},\frac{1}{3}\ket{\sigma^{-}_{\mathrm{X}}}\bra{\sigma^{-}_{\mathrm{X}}},\dots \frac{1}{3} \ket{\sigma^{-}_{\mathrm{Z}}}\bra{\sigma^{-}_{\mathrm{Z}}}\right\}, \label{eq:app:POVMPauli}
\end{align}
which corresponds to the projections on the eigenstates of the Pauli matrices. Note that the normalization factor $3$ corresponds to the total number of different Pauli basis measurement settings. If we implement single-qubit Pauli basis measurements on larger systems, the set of tomographically complete measurements contains all tensor products of the states in Eq.~(\ref{eq:app:POVMPauli}). 
For the two-qubit system we investigate the main text, we obtain the following $36$ operators
\begin{align}
   \operatorname{P}_i= \Bigg\{&\frac{1}{9}\ket{\sigma^{+}_{\mathrm{X}}}\bra{\sigma^{+}_{\mathrm{X}}}\otimes \ket{\sigma^{+}_{\mathrm{X}}}\bra{\sigma^{+}_{\mathrm{X}}},
   \dots, \nonumber \\
   &\frac{1}{9}\ket{\sigma^{-}_{\mathrm{Z}}}\bra{\sigma^{-}_{\mathrm{Z}}} \otimes \ket{\sigma^{-}_{\mathrm{Z}}}\bra{\sigma^{-}_{\mathrm{Z}}}\Bigg\}, \label{eq:app:POVMPauliTwoqubits} 
\end{align}
where the prefactor of $9$ corresponds again to the number of distinct Pauli basis measurement settings.

\subsection{Unbiased estimators}
In general, the least-squares estimator $\rhoLS$ is obtained from the inversion of the linear map \mbox{$\mathcal{A}:\mathcal{L}(\mathbb{H}_d) \mapsto \mathbb{R}^m$}, which has been induced by Born's rule Eq.~(\ref{Bornsrule}). If the map $\mathcal{A}$ is injective, it can be inverted as follows:
\begin{equation}
    \rhoLS=\left(\mathcal{A}^{\dagger} \mathcal{A}\right)^{-1}\left(\mathcal{A}^{\dagger}\left(f_N\right)\right), \label{linInvEst}
\end{equation}
where $f_N$ are the observed frequencies. The actual expression for Eq.~(\ref{linInvEst}) depends on the operators $P_i$, and in the following we discuss the case of single-qubit Pauli basis measurements. 

\subsubsection{Linear inversion for single-qubit Pauli measurements}
In this section we use the convention $(1,\operatorname{Tr(\sigma_{\operatorname{X}}\rho)},\operatorname{Tr(\sigma_{\operatorname{Y}}\rho)},\operatorname{Tr(\sigma_{\operatorname{Z}}\rho)})$ to parametrize the Bloch vector as vector in $\mathbb{R}^4$. By this convention we mean to express states as linear combination of the identity and the Pauli matrices, whereby for the latter their expectation values serve as coefficients. We represent the linear map $\mathcal{A}$ as matrix $\operatorname{A}$ by taking into account our convention for vectors:
\begin{align}
    \operatorname{A}=\frac{1}{6}\left(\begin{array}{cccc}
1 & 1 & 0 & 0 \\
1 & -1 & 0 & 0 \\
1 & 0 & 1 & 0 \\
1 & 0 & -1 & 0 \\
1 & 0 & 0 & +1 \\
1 & 0 & 0 & -1
\end{array}\right). \label{eq:app:matrixLinMap}
\end{align}
We use the Moore-Penrose pseudo inverse~\cite{moore_1920,penrose_1955} to invert the linear map induced by Born's rule, and thus, the least-squares estimator for local Pauli measurements of a single-qubit is given as follows 
\begin{align}
    \rhoLS &=\left(\operatorname{A}^T \operatorname{A}\right)^{-1} \operatorname{A}^T   f \nonumber \\
    &= \frac{1}{2}\left(\begin{array}{cccccc}
1/3 & 1/3 & 1/3 & 1/3 & 1/3 & 1/3 \\
1 & -1 & 0 & 0 & 0 & 0 \\
0 & 0 & 1 & -1 & 0 & 0 \\
0 & 0 & 0 & 0 & 1 & -1
\end{array}\right) \left(\begin{array}{c}
f_0  \\
f_1  \\
f_2  \\
f_3 \\
f_4\\
f_5
\end{array}\right) \nonumber \\
&= \frac{1}{2} \left(\begin{array}{cccc}
1, & f_1-f_2, & f_3-f_4, & f_5-f_6
\end{array}\right)^T,
\label{eq:app:linOpt}
\end{align}
where the vector contains the observed frequencies $f_i$ corresponding to the set of measurements in Eq.~(\ref{eq:app:POVMPauli}). In the last line we use the fact that the individual probabilities for each Pauli basis setting sum up to one, e.g. $(f_1+f_2=1)$. Thus, if we rewrite the vector in Eq.~(\ref{eq:app:linOpt}) in terms of the Pauli matrices, we find that the least-squares estimator for local, single-qubit Pauli basis measurements corresponds to a linear combination of the all Pauli matrices: 
\begin{align}
    \rhoLS = \frac{1}{2} \left(\openone+ (f_1-f_2)\sigma_{\operatorname{X}}+(f_3-f_4)\sigma_{\operatorname{Y}} +(f_5-f_6) \sigma_{\operatorname{Z}} \right),
\label{eq:app:linInv}
\end{align}
where we identity the differences in the probabilities as the components of the Bloch vector. Note that the estimated state $\rhoLin$ has unit trace due to construction of the measurement set $P_i$, but it is not guaranteed to be positive semidefinite for every configuration of the outcome frequencies $f_N$. The $n$-qubit least-square matrix  $A_{\mathrm{n}}$ is obtained from the $n$-fold tensor product of the single-qubit matrix $A$ in Eq.~(\ref{eq:app:linOpt}).

\subsection{Biased estimators}
If one wishes to ensure that the estimated quantum state $\rhoBiased$ is positive semidefinite and has unit trace, then the used estimator is necessarily biased \cite{SchwemmerPRL2015}. Here, we discuss our convex optimization approach of the main text, which is related to the work of Smolin \ea~\cite{smolin2012} and Guta \ea~\cite{Guta_2020}, because they introduce physical state estimators as optimization problems and we follow their spirit. Furthermore, we briefly review the maximum-likelihood estimator, because it is a commonly used method.
\subsubsection{Convex optimization estimator}
As we have proposed in the main text, the state $\rhoBiased$ is obtained from the convex optimization problem of finding the closest physical state $\rhoBiased$ to the least-squares estimate $\rhoLS$ in the terms of the squared Frobenius norm: 
\begin{align}
    \rhoBiased &= \argmin _{\substack{\operatorname{Tr}(\varphi) = 1 \\ \varphi \succcurlyeq 0}} ||\rhoLS-\varphi||^{2}_2 = \argmin _{\substack{\operatorname{Tr}(\varphi) = 1 \\ \varphi \succcurlyeq 0}} \trace \left( \left(\rhoLS-\varphi\right)^2 \right), \nonumber \\
    \label{eq:app:convOptimProblem}
\end{align}
where we used the fact that we are considering Hermitian matrices to simplify the expression given by the Frobenius norm. 

Here, we shortly discuss the analytical solution to the convex optimization problem, Eq.~(\ref{eq:app:convOptimProblem}), for a single qubit. Let $\rhoLS$ and $\rhoBiased$ be the unbiased and biased single-qubit estimate to a single-qubit quantum tomography experiment, respectively. The positivity constraints in Eq.~(\ref{eq:app:convOptimProblem}) reduce to a single constraint for the single-qubit case, which is that the purity of the quantum state has to be smaller equal than one to be a physical state. If the unbiased estimate lies beyond the Bloch sphere, the closest biased estimate lies on the surface of the Bloch sphere, and is obtained by projecting the unbiased estimate on the surface of the Bloch sphere. Thus, the analytical solution for the single-qubit biased estimate is:
\begin{align}
    \rhoBiased =\left\{\begin{array}{ll}
\rhoLS & \text { if } \trace\left(\rhoLS^2\right) \leq 1 \\
\frac{1}{2}\left(\openone + \frac{1}{u}\sum^{3}_{i=1} u_i \sigma_i\right) & \text { else } 
\end{array}\right.,
\end{align}
where $u_i = \trace\left(\sigma_i \, \rhoNonPhys\right)$ is the experimentally obtained Bloch vector component corresponding to the $i$-th Pauli basis measurement, and the normalization coefficient \mbox{$u=\sqrt{u^2_{\operatorname{X}} + u^2_{\operatorname{Y}} + u^2_{\operatorname{Z}}}$} is the length of the Bloch vector of the unbiased estimate.

We are solving Eq.~(\ref{eq:app:convOptimProblem}) for more than one qubit numerically by using the convex optimization tool 'cvxpy'~\cite{diamond2016cvxpy,agrawal2018rewriting}.

\subsubsection{Maximum-likelihood estimator}\label{app:sec:MLE}
Here, we want to review another prominent biased estimator, the maximum-likelihood estimator (MLE)~\cite{HradilPRA1997}, which estimates the quantum state $\rho_{\text{MLE}}$ maximizing the probability of observing the measurement outcome frequencies:
\begin{align}
   \rho_{\text{MLE}} = \argmax_{\substack{\operatorname{Tr}(\rho) = 1 \\ \rho \succcurlyeq 0}}\mathcal{L}(\rho) = \argmax_{\substack{\operatorname{Tr}(\rho) = 1 \\ \rho \succcurlyeq 0}} \prod_i^m \trace \left(\operatorname{P}_i \rho \right),\label{eq:app:MLECore}
\end{align}
where $\mathcal{L}$ is a non-linear functional in $\rho$ modeling the probability distribution underlying the measurement, we assume for the explicit implementation of $\mathcal{L}$ that the outcomes corresponding to $\trace \left(\operatorname{P}_i \rho \right)$ are independently and identically distributed random variables. We refer the reader to the Ref.~\cite{Blume-Kohout_2010} or Ref.~\cite{numericalRecipiesC} for a detailed discussion on the difference between least-squares estimator and the MLE.

A critical issue of our proposed estimator and MLE is that it predicts quantum states with zero eigenvalues, as pointed out in Ref.~\cite{Blume-Kohout_2010}, and this causes an incompatibility with error bars. Furthermore, the conclusion that an estimated outcome will never appear for future measurement execution is a very strong claim, in particular if the reasoning is based on finite measurement statistics. To circumvent this, Refs.~\cite{HradilPRA1997,ALTEPETER2005105} suggest computing many estimates $\rho_{\text{MLE}}$ by simulating measurement frequency data by sampling from a Poissonian distribution with the mean values being the observed frequencies, which has been criticized in Ref.~\cite{Blume-Kohout_2010}.

\section{Necessary purity to detect systematic errors}\label{app:criticalPurity}
In this section, we support the discussion on the smallest necessary purity of the true, underlying quantum state $\rho$ from the main text. We assume having local systematic errors in the implementation of the measurement basis, which is modeled by the misalignment matrix $M$ from the main text, and $M$ has to have normalized normalized rows in order to represent a physical measurement. In particular, the normalization of the rows has to be such that the eigenvalues of each misaligned matrix $\tilde{\sigma}_i$ are $\pm 1$. For example, if the misaligned Pauli Z basis is \mbox{$\tilde{\sigma}_{\operatorname{Z}}=\alpha \, \sigma_{\operatorname{Y}} + (1-\alpha)\, \sigma_{\operatorname{Z}}$}, the normalization $\mathcal{N}$ is determined from \mbox{$(\alpha^2+(1-\alpha)^2)/\mathcal{N}=1$}.

\subsection{Single-qubit system} \label{sect:singleQubitPurity}
We mention here the details on the Bloch components that we only briefly mention in the main text. For a single-qubit error $M$, the erroneous Bloch vector is given by $\Tilde{\bold{u}} = \operatorname{M} \bold{u}$, because each Bloch vector component $u_i$ transforms as follows:
\begin{align}
    \Tilde{u}_i = & \trace\left( \Tilde{\sigma}_i \rho \right) = \trace \left( \sum_{j} \operatorname{M}_{ij} \sigma_{j}  \rho \right) = \sum_{j}  \operatorname{M}_{ij} \trace \left( \sigma_{j} \rho \right) \nonumber \\ =& \sum_{j} \operatorname{M}_{ij} u_{j} .  \label{eq:BlochVector}
\end{align}
The erroneous Bloch vector components $\Tilde{u}_k$ determine the erroneous least-squares estimator $\tilde{\rho}_{\text{LS}}$:
\begin{align}
\Tilde{\rho}_{\text{LS}} = \frac{1}{2}\left(\openone+ \sum_{k, l} \trace(\sigma_l\rho) \operatorname{M}_{kl} \sigma_k\right) =\frac{1}{2}\left(\openone+ \sum_{k} \Tilde{u}_k \sigma_k\right). \label{eq:linInvBloch}
\end{align}
In order to determine the minimal, necessary state purity for detecting a systematic error we need the purity of the erroneous least-square estimator:
\begin{align}
    \trace \left(\Tilde{\rho}_{\text{LS}}^2\right)=& \frac{1}{4}\left(2+\trace \left(\sum_{k, k^{\prime}} \trace(\tilde{\sigma}_{k}\rho) \sigma_k \trace(\tilde{\sigma}_{{k^{\prime}}}\rho) \sigma_{k^{\prime}} \right)\right)\nonumber \\=&\frac{1}{2}\left(1+\sum_{k}\left( \trace(\tilde{\sigma}_{k}\rho)\right)^2\right)\nonumber\\=&\frac{1}{2}\left(1+\sum_{k} \Tilde{u}_k^2 \right) = \frac{1}{2}\left(1+ || \Tilde{\bold{u}} ||^2\right). \label{eq:errorPurity}
\end{align}
Note that the purity of the single-qubit state without errors corresponds to Eq.~(\ref{eq:errorPurity}), if we exchange the erroneous Bloch vector $\tilde{\bold{u}}$ by the correct one $\bold{u}$.

\subsection{Two-qubit system}
As in the main text, we have local, systematic errors $M_1$ and $M_2$ on the first and second qubit, and no correlated error. From these misalignment matrices we obtain the erroneous local Bloch vectors, \mbox{$\Tilde{\bold{u}}=\operatorname{M}_{1}\bold{u}$} and \mbox{$\Tilde{\bold{v}}=\operatorname{M}_{2}\bold{v}$}, and the erroneous correlation matrix is given as \mbox{$\Tilde{\operatorname{R}}=\operatorname{M}_{1}\operatorname{R}\operatorname{\,M}^T_{2}$}, because its components transform as follows:
\begin{align}
\tilde{r}_{ij}&=\trace\left(\tilde{\sigma_i} \otimes \tilde{\sigma}_j \, \rho\right)=\trace \left(\left(\sum_{l,m} \operatorname{M}_{1,il} \sigma_l \otimes \operatorname{M}_{2,jm} \sigma_m\right)  \rho\right) \nonumber \\ &= \trace \left(\left(\sum_{l,m} \operatorname{M}_{1,il} \sigma_l \otimes  \sigma_m \operatorname{M}^T_{2,mj}\right)  \rho\right) \nonumber \\ &= \sum_{l,m} \operatorname{M}_{1,il} \trace \left(\left( \sigma_l \otimes  \sigma_m \right)  \rho\right) \operatorname{M}^T_{2,mj} \nonumber \\ &= \sum_{l,m} \operatorname{M}_{1,il} r_{lm} \operatorname{M}^T_{2,mj}.
\end{align}
We use the erroneous least-squares estimator as an unbiased estimator, which we express in the tensor product basis of Pauli matrices, $\sigma_{\mu} \otimes \sigma_{v}$, with erroneous components $\tilde{r}$:
\begin{equation}
    \tilde{\rho}_{\text{LS}} =\frac{1}{4} \tilde{r}_{\mu v}\, \sigma_{\mu} \otimes \sigma_{v}.
\end{equation}

\subsubsection{Systematic errors detection with separable states}
\label{app:sepStatespurity}
Here we discuss how we modified the optimization problems of the main text to investigate the difference between entangled and product states in their capability to detect specific errors. As discussed in the main text, the PPT criterion~\cite{PeresPRL_1996_PPT,HORODECKI_PLA_1996} utilizes the partial transpose on a bipartite qubit system that is defined by:
\begin{equation}
    \rho^{\operatorname{T}_2} := (\openone\otimes \operatorname{T}) (\rho),
\end{equation}
where $\operatorname{T}_2$ denotes the map corresponding to the transposition of the subspace of qubit two.

We can directly insert $\rho^{\operatorname{T}_2}$ into the positivity conditions, presented in the main text, to ensure the positivity of the partial transpose of $\rho^{\operatorname{T}_2}$. To reduce the number of positivity constraints of the partial transposed state, $\zeta(\rho^{\operatorname{T}_2})$, we investigate how the terms of positivity conditions, presented in the main text, change under partial transposition on the second qubit. Only the Pauli $Y$ matrix changes the sign under partial transposition, \mbox{$\operatorname{T}(\sigma_{\operatorname{Y}})\mapsto - \sigma_{\operatorname{Y}}$}. Additionally, separable states remain separable (and keep their purity) under local unitary transformations. Since one can flip the sign of two Paulis (such as $\sigma_{\operatorname{X}}$ and $\sigma_{\operatorname{Z}}$) simultaneously by a local unitary transformation, one may interpret the partial transposition (assisted by a local unitary transformation) so that the entire local Bloch vector of the second qubit $\bold{v}$ changes the overall sign, $\operatorname{T}(\bold{v}) \mapsto -\bold{v}$. 
In this way, transposition changes the sign of all terms $r_{i,j}$ in the correlation matrix $\operatorname{R}$,
\begin{equation}
    \operatorname{T}_2(\operatorname{R}) = \left[\begin{array}{lll}
    -r_{11} & -r_{12} & -r_{13} \\
 -r_{21} & -r_{22} & -r_{23} \\
 -r_{31} & -r_{32} & -r_{33}
\end{array}\right].
\end{equation}
The changes of the local Bloch vector of qubit two, $\bold{v}$, and the correlation matrix $\operatorname{R}$ under partial transposition on the system of the second qubit affect only the sign in three terms of the positivity conditions of $\zeta_2$ and $\zeta_3$, which we highlight in the following equation in red. We denote the two conditions for the positivity of the partial transpose $\varepsilon_2$ and $\varepsilon_3$:
\begin{align}
 &\varepsilon_2(\rho)=  \left(\|\stackrel{\leftrightarrow}{r}\|^2-2\right) - 2\left(\bold{u}^{\dagger} \operatorname{R} \bold{v}+\colorbox{red!40}{$\operatorname{det} \operatorname{R}$}\right) \leq 0 \\
 &{\varepsilon}_3(\rho)=  -8\left(\bold{u}^{\dagger} \operatorname{R} \bold{v}+\colorbox{red!40}{$\operatorname{det} \operatorname{R}$}\right)-\left(\|\stackrel{\leftrightarrow}{r}\|^2-2\right)^2\nonumber \\ &\phantom{A}+4\left(\|\bold{u}\|^2 \, \|\bold{v}\|^2+\left\|\bold{u}^{\dagger} \operatorname{R}\right\|^2+\|\operatorname{R} \bold{v}\|^2+\|\operatorname{R}_{\text{cof}}\|^2\right) \nonumber \\
 &\phantom{A}+\colorbox{red!40}{$8 \bold{u}^{\dagger} \operatorname{R}_{\text{cof}} \bold{v}$} \leq 0. \label{app:eq:2qubitzetacondition-2}
\end{align}

\subsection{Two-qubit results for necessary purity to detect systematic errors}
\label{sec-app-entanglement}
In this section we present the results supporting the key statements of the main text. We show that the minimum necessary state purity differs for two- and single-qubit states when having the same local systematic error on the first qubit. Finally, we discuss and show results for local single-qubit errors that can only be detected with entangled states.

\subsubsection{Single-qubit vs. two-qubit states}
\begin{figure}
    \centering
     \includegraphics{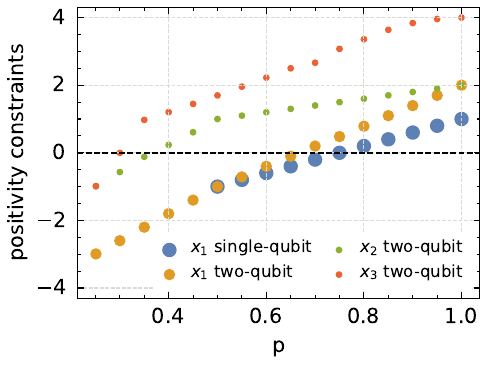}  
    \caption{The maximal reachable positivity constraints $x_l$, as stated in the main text, of the erroneous state are presented as a function of the single- and two-qubit quantum state purity $p$. We assume the systematic error of measuring $\PauliY$ instead of $\PauliZ$ on one qubit. The black dashed line is located at a value of zero, and if a condition is $x_l$ is above that line, it means that there exists a quantum state with the respective purity which can detect the error. The minimal necessary state purity is $0.75$ and $0.3$ for single- and two-qubit states, respectively.}
    \label{fig:numResPur}
\end{figure}

Here, we consider again the single-qubit example as from the main text, where we assume to have the error of measuring $\sigma_{\operatorname{Y}}$ instead of $\sigma_{\operatorname{Z}}$ on the first qubit. We compare the error detection capabilities of single- and two-qubit states. In the case of two qubits, we numerically determine the values $x_l$ by optimizing the three problems, stated in the main text, for purity values of $\rho$ ranging from $0.25$ to $1$. In the case of a single qubit, we optimize $x_1$ of the main text for purity values of $\rho$ ranging from $0.5$ to $1$. Fig.~\ref{fig:numResPur} shows the results for the single- and two-qubit optimization $x_l$ as function of the state purity $p$, where the blue dots are the results from the single-qubit optimization $x_1$, and the orange, green and red dots correspond to the two-qubit results of $x_1$, $x_2$ and $x_3$, respectively. Note that we analytically determine the value of $x_1$ for the lowest single-qubit purity $0.5$ of the single-qubit state, to avoid numerical problems with optimization at this edge point.

These results show that the minimum necessary purity for two-qubit states is significantly smaller than that for single-qubit states. In particular, the results show that two-qubit quantum states are more sensitive to detect an error in a single-qubit Pauli measurement, due to the additional $\zeta_2$ and $\zeta_3$ positivity constraints, which are violated already at a purity of $0.3$ and $0.4$, respectively. The single- and two-qubit $\zeta_1$ inequalities are violated at similar purity values of $0.75$ and $0.7$, respectively, which agrees with the minimal necessary purity of $0.75$ in Fig.~\textcolor{blue}{1c}, where the distance $D$ vanishes for states with a purity below $0.75$. Of course, one has to keep in mind that the purities of quantum states with a different number of qubits are sometimes difficult to compare.

\subsubsection{Tensor products with single-qubit state copies}
We have also studied whether one can enhance the single-qubit error detection capabilities by adding uncorrelated ancillas. We again consider the error of measuring $\sigma_{\operatorname{Y}}$ instead of $\sigma_{\operatorname{Z}}$ on the first qubit. We use the true underlying single-qubit state $\rho^{s}_{0.7}$ with purity of $0.7$, which maximizes $x_1$ of the main text. In particular, we consider three cases of tensor products. First, we find that the tensor product of an erroneous state copy $\tilde{\rho}^{s}_{0.7}$ with a correct copy $\rho^{s}_{0.7}$ does not violate any positivity condition. Second, the same holds for a tensor product of two copies of the erroneous state $\tilde{\rho}^{s}_{0.7}$. Finally, if we consider $\tilde{\rho}^{s}_{0.7}$ together with any eigenvector corresponding to the positive eigenvalue of $\sigma_i$, we only get a maximal value of $0$ for $\zeta_2$ and $\zeta_3$. Table~\ref{tab:prodStates} summarizes the results for $\zeta_i$ depending on the type of product state. In summary, these attempts to enhance the detection capability of a single-qubit error by employing some tensor products in different cases have not been successful. 

\begin{table}[b]
\caption{\label{tab:prodStates} Single-qubit error ($\PauliY$ measurement for $\PauliZ$) detection with several product states containing the erroneous single-qubit state $\tilde{\rho}^{s}_{0.7}$ with purity $0.7$.}
\begin{ruledtabular}
\begin{tabular}{lccc}
product states&$\zeta_1$&$\zeta_2$&$\zeta_3$\\
\colrule
 $\tilde{\rho}^{s}_{0.7}\otimes \openone$ \textrm{\footnote{corresponds to single-qubit case}} & -0.20 & - &-\\
 $\tilde{\rho}^{s}_{0.7}\otimes \rho^{s}_{0.7}$\textrm{\footnote{true, underlying state}} & -1.480 & -0.120 &-0.014\\
 $\tilde{\rho}^{s}_{0.7}\otimes \tilde{\rho}^{s}_{0.7}$ & -0.760 & -0.040 &-0.002\\
$\tilde{\rho}^{s}_{0.7}\otimes (1,0,0)^{\operatorname{T}}$\textrm{\footnote{same results are obtained for the other eigenstates $(0,1,0)^{\operatorname{T}}$ and $(0,0,1)^{\operatorname{T}}$}} & -0.400 & 0 &0\\
\end{tabular}
\end{ruledtabular}
\end{table}

\subsubsection{Error detection capability: entangled vs. separable states}\label{app:entanglementUseful}
In this section we summarize the difference between product and entangled two-qubit states in their capability to detect local systematic errors on the first qubit. We focus on the case of a severe systematic error on a single qubit, as this is the most probable error. As stated in the main text, this error can be described by the matrix $\operatorname{M}_1(\alpha,\beta,\gamma,\delta)$, and we assume that there is no systematic error present on the second qubit, i.e. $\operatorname{M}_2=\openone$. 

We checked all permutations of the four angles for the interval $\{0,\frac{\pi}{8},\frac{\pi}{4},\frac{3\pi}{8},\frac{\pi}{2}\}$. For each angular setting we determine the values $x_l$ for the whole state space and for separable states as stated in the main text, where we vary the purity values from $0.32$ to $0.38$. The optimization over the whole state space includes both entangled and separable states, whereas the optimization over the separable states includes only those, which enables us to compare entangled and separable states in their error detection capability. 

We have checked $625$ different angular settings, where $25$ of those angular settings correspond to not having an error, because the $\sin{(0)}$ terms let $\operatorname{M}_1$ result in the identity matrix. In summary, we have $600$ erroneous angular settings. 

We find that the optimization problem resulting in $x_3$ is the most stringent condition, because it is already positive at the smallest purity values. We observe this property of $x_3$ for all erroneous settings and for both entangled and separable states. Furthermore, we find that the first positive value of conditions $x_1$ and $x_2$ strongly depends on the angular setting $\operatorname{M}_1$. However, the first positive value of $x_3$ is almost the same for all angular settings, which we find in the interval between $0.29$ and $0.34$, and $54\%$ of all erroneous settings return the smallest necessary state purity of $0.31$. 

For most of the cases, we find that the minimum necessary state purity is the same for entangled and separable states. In some of these cases, the entangled states return larger $x_l$ values than separable states, but this happens only at purity values above the minimal necessary state purity, which has been the same for both state types.

\begin{figure*}
    \centering
    \includegraphics{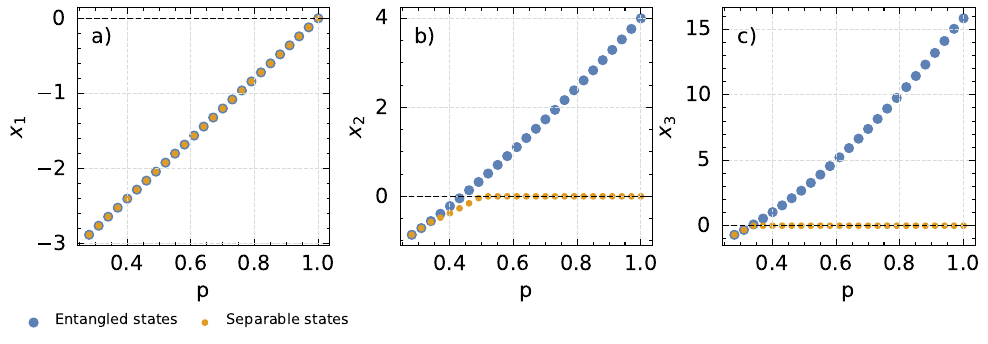}
    \caption{Numerical results for the maximal values of $x_1$ in a), $x_2$ in b) and $x_3$ in c) depending on the purity of the true underlying two-qubit state $\rho$ for the error of interpreting the $\operatorname{Y}$ and $\operatorname{Z}$ as $\operatorname{Z}$ and $\operatorname{Y}$ Pauli measurement, respectively. The blue and orange dots represent the $x_l$ values for entangled and separable states as discussed in the main text. If a point is above the black line at $x_l=0$, it means that one of the positivity constraints is violated, and thus, there exists at least one state that can detect the error. Only for entangled states, the results of $x_2$ and $x_3$ are above the dashed line at a purity of $0.45$ and $0.35$, respectively. This error relates to the partial transposition on a subsystem, which explains why only entangled states are sensitive to this error.
    }
    \label{fig:EntVSSEP}
\end{figure*}

The error $\operatorname{M}_1$ with $\alpha=\gamma=\frac{\pi}{2}, \beta=\delta=0$, is the only one for which the results of entangled and product states differ drastically, and corresponds to the Pauli basis permutation: $\operatorname{X}\mapsto \operatorname{X}, \operatorname{Y} \mapsto \operatorname{Z}, \operatorname{Z} \mapsto \operatorname{Y}$. Figs.~\ref{fig:EntVSSEP}\textcolor{blue}{a-c} show the results $x_l$ of the optimization problems for entangled and separable states as blue and orange dots, respectively. The black dashed line is at $x_l=0$, and if a point is above the line, it means that there exists a state that can detect the error. We find that only entangled states can detect this error, as seen in Figs.~\ref{fig:EntVSSEP}\textcolor{blue}{b\&c}, where only the blue dots are above the black, dashed line at critical purities of $0.45$ and $0.35$, respectively. The product state results, orange dots, are all below the dashed line, which means that they cannot detect the error.

\begin{figure*}
    \centering
    \includegraphics{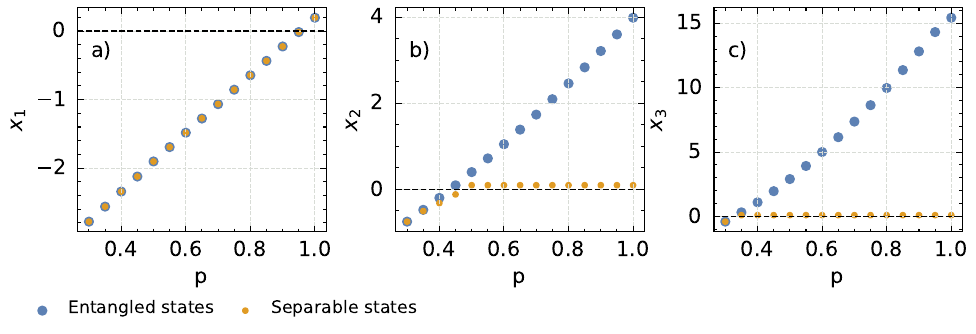}
    \caption{Numerical results for the maximal values of $x_1$ in a), $x_2$ in b) and $x_3$ in c) depending on the purity $p$ of the true underlying two-qubit state $\rho$ for the error $\operatorname{M}_1$ with angles $\alpha=\gamma = \frac{\pi}{2}+\eta$ and $\beta = \delta = \eta$. This error corresponds to exchanging the $\operatorname{Y}$ and $\operatorname{Z}$ basis, but with a small angular offset of $\eta=\SI{0.05}{\radian} = \SI{2.9}{\degree}$. The blue and orange dots represent the $x_l$ values for entangled and separable states, discussed in the main text. The black dashed line is at $x_l=0$, and marks the threshold value when the positivity constraint is violated and at least one state exists, which can detect the systematic error. The violations that can be reached with separable states are much smaller than the ones for entangled states, therefore entangled states can help in detecting this kind of error.}
    \label{fig:robustness}
\end{figure*}

The above-mentioned error of permuting two bases corresponds to the partial transpose map together with an appropriate local unitary transformation. Thus, it is understandable that separable states are unable to detect those errors. From the above discussion, the following questions arise naturally: what happens if the adjustment angles are close to (but not exactly equal to) the partial transpose error? and is there still a difference between the entangled and separable states? To address this questions, we study the error mentioned above with a small angular offset $\eta$: $\alpha=\gamma = \frac{\pi}{2}+\eta$ and $\beta = \delta = \eta$. We find for a value of $\eta=\SI{0.05}{\radian} = \SI{2.9}{\degree}$ that again only entangled states violate the conditions $x_2$ and $x_3$ significantly as one can see in Fig.~\ref{fig:robustness}\textcolor{blue}{b-c}. For separable states, there is a violation, but it is much smaller and difficult to characterize experimentally. In this sense, entangled states can help detect systematic errors.

To conclude this section, we find that entangled states are more powerful than separable states in detecting errors on one qubit, as our discussion shows. Additionally, we find that entangled and separable states have the same minimal necessary state purity for many errors. However, in some cases, the numerical value of $x_l$ is larger for entangled states than for separable states, but this happens only for state purities beyond the smallest necessary state purity.

\section{Statistical and systematic error distinction\label{lab:statisticalMeaning}}
In this section we utilize the vector Bernstein inequality from Ref.~\cite{degois2023userfriendly} to state a probability $\delta_{\text{sta}}$ that the distance $D$ between the least-squares and biased estimate, Eq.~(\textcolor{blue}{3}), is caused by statistical errors. In order to be able to use the vector Bernstein inequality, we have to reformulate the distance $D$ such that it contains the true, underlying quantum state $\rho$. The distance $D$ is the smallest distance between the least-squares estimator $\rhoLS$ and the physical state space, and thus, it is strictly smaller than the distance between the least-squares estimate and the true, underlying quantum state $\rho$: \mbox{$D=\left \| \rhoLS-\rhoBiased \right\|_{2} \leq \left\| \rhoLS-\rho \right\|_{2}$}. Furthermore, we interpret quantum state tomography as sum of independent, zero-mean, vector-valued random variables, where we define $\sigma$ as the average of the variances of the random variables. With this assumptions we utilize Eq.~(A16) from Ref.~\cite{degois2023userfriendly}, which makes us of the pseudo inverse map induced by Born's rule ($\mathcal{A}: \rho \mapsto f$). In summary, we obtain the following bound on the probability $\delta_{\text{sta}}$ to observe statistical errors:
\begin{align}
 \delta_{\text{sta}} &=\,\mathbb{P}\left[\left\|\rhoLS-\rhoBiased\right\|_2 \geqslant \tau\right] \leq \mathbb{P}\left[\left\|\rhoLS-\rho \right\|_2 \geqslant \tau\right] \nonumber \\ &\leq 8 \exp \left[-\frac{N \tau^2}{2 \sigma^2} \frac{3}{3+\sqrt{2} \tau / \sigma}\right],
\end{align}
where $\tau$ is a finite distance. For local Pauli measurements on $n$ qubits $\sigma$ is equal to $5^{n/2}$.

\section{Experimental methods\label{app:dataEvaluation}}
\subsection{Photon source and experimental setup}

\begin{figure*}
    \centering
    \includegraphics[width=\linewidth]{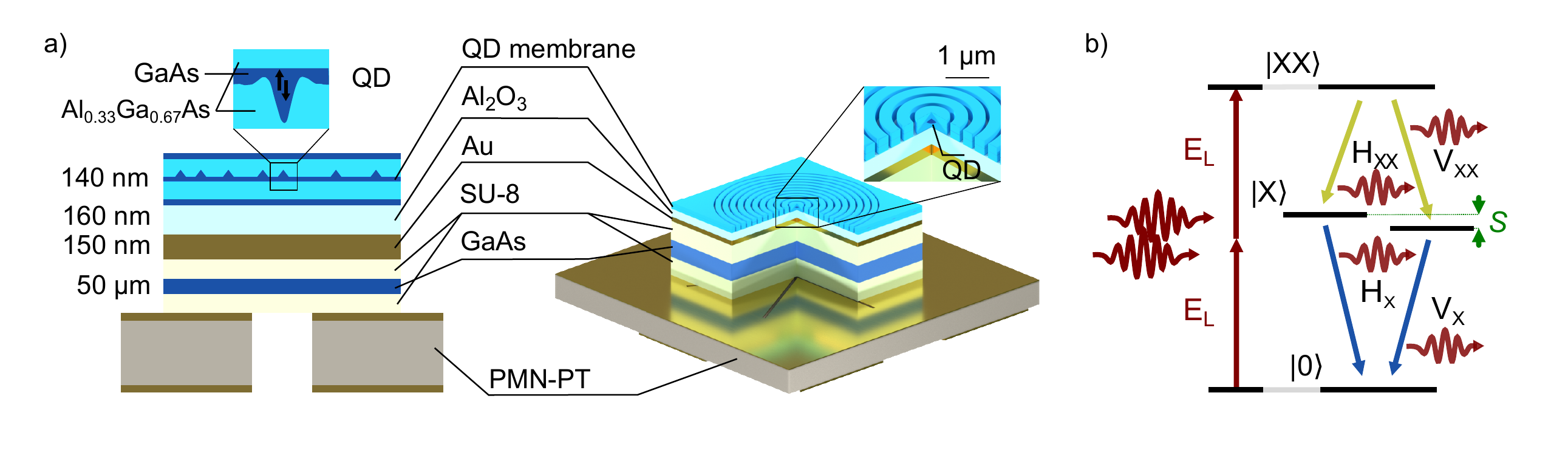}
    \caption{a) Sketch of the entangled photon source. A layer with GaAs QDs is fabricated via droplet etching epitaxy at the middle of a \( 140 \,\text{nm} \) thick $\text{Al}_{0.33}\text{Ga}_{0.67}\text{As}$ membrane. The membrane is integrated on top of a micromachined piezoelectric substrate to induce controlled strain in the material. A circular Bragg grating microstructure is etched around a single QD to improve photon collection from above. b) Energy level picture of the biexciton-exciton radiative cascade that generates polarization-entangled photons. A fine structure splitting $S$ separates the two bright exciton states. A resonant two-photon absorption process directly populates the biexciton state.
    }
    \label{fig:qd-basics}
\end{figure*}

A single GaAs/AlGaAs quantum dot (QD) embedded in a circular Bragg resonator cavity, also known as a bullseye cavity, serves as the source of entangled photons. The QD is a semiconductor nanostructure formed by tens of thousands of atoms, whose electronic structure features discrete energy levels due to quantum confinement. This is a distinctive feature of quantum emitters, namely light sources that rely on deterministic emission mechanisms and can produce photonic states with small multiphoton components~\cite{Hanschke2018}. The QD is fabricated by Al-droplet etching epitaxy, a technique that creates symmetric conical nanoholes in an $\text{Al}_{0.33}\text{Ga}_{0.67}\text{As}$ matrix, here about $\SI{7}{\nano\meter}$ in height and $\SI{50}{\nano\meter}$ in base diameter~\cite{huo2013ultra}, that are filled with GaAs. The quantum dot is embedded in a photonic microcavity, here a circular Bragg resonator, that enhances photon extraction. The whole structure is sketched in Fig.~\ref{fig:qd-basics}\textcolor{blue}{a}. More details on source fabrication and further micro-processing steps, in addition to the full sample design and optical performance, are described in Ref.~\cite{rota2024source}. 

Pairs of polarization-entangled photons are generated by the physical mechanism of the biexciton-exciton cascade, which is depicted in Fig.~\ref{fig:qd-basics}\textcolor{blue}{b}. If two electron-hole pairs are confined in the quantum dot, a so-called biexciton state with total angular momentum $J=0$ along the quantization axis defined by the growth direction, is created. This excited state decays via spontaneous emission, a process that can follow two radiative paths based on the presence of two possible intermediate states, namely bright exciton states with different projections of the total angular momentum along the main confinement axis (superposition of $\operatorname{M}_z = \pm 1$ states). The energy of the first optical transition (biexciton to exciton) is lower than the energy of the second transition (exciton to ground state) due to differences in Coulomb interaction in the excitonic states. Following the relevant optical selection rules, when the two decay paths are indistinguishable, that is when the two intermediate bright exciton states are degenerate, the two photons of the radiative cascade are emitted in the $\ket{\phi^{+}} = 1/ \sqrt{2} (\ket{HH} + \ket{VV})$ state of polarization~\cite{BensonPRL2000}.  

Including non-idealities, the two-qubit density matrix that describes the polarization state of the two correlated photons depends on a few parameters (see also Appendix~\ref{app:rotationPaulibasis}). The main factor that influences the density matrix is the fine structure splitting, that is the energy splitting between the two intermediate bright-exciton states of the radiative cascade.

Fine-structure splitting introduces a phase in the entangled state that depends on the randomly distributed emission time of the exciton-to-ground state transition~\cite{Hudson2007PRL,RotaIEEE2020}. Under the assumption of no spin dephasing processes, the state follows the following form: $\ket{\psi} = \frac{1}{\sqrt{2}} \left( \ket{H H} + e^{i E_{\text{FSS}} t / \hbar} \ket{V V} \right)$, where $E_{\text{FSS}}$ is the fine structure splitting and $t$ is the emission time of the exciton, where zero is the emission time of the biexciton. If the emission is time gated and the detector resolution is fast enough, this phase might be resolved, however, otherwise the phase evolution results in a mixed density matrix, which deviates from the dominant term $\ket{\phi^+}= \frac{1}{\sqrt{2}} \left( \ket{HH} + \ket{VV} \right)$. The amount of mixedness depends on the ratio between fine structure splitting and exciton recombination lifetime. Without the assumption of no spin dephasing, the state evolution can be found in matrix form in Appendix~\ref{app:furtherErrors}. Here we control the fine structure splitting to modify the density matrix of the emitted two-photon states at a source level. Our entangled photon source is integrated on top of a micromachined piezoelectric substrate which allows us to induce a controlled strain and finely tune the value of fine structure splitting. In this experiment, the QD source is operated at three different levels of fine structure splitting, namely $\SI{0.6}{\micro \electronvolt}$ (negligible effect on the entanglement), $\SI{7.9}{\micro \electronvolt}$, and $\SI{13.9}{\micro \electronvolt}$. These three values of fine structure splitting correspond to a state purity of $0.92$, $0.65$, and $0.56$, in order. The lifetime is $\SI{93}{\pico \second}$ and $\SI{42}{\pico \second}$ for the exciton-to-ground-state and biexciton-to-exciton transition, respectively.

The photon source is cooled to $\SI{5}{\kelvin}$ inside a low-vibration closed-cycle He cryostat, where a single QD is optically excited by focusing laser light through a $0.81$ NA objective positioned within the cryostat in a confocal arrangement. The radiative cascade is prompted by populating the biexciton state of the QD via two-photon resonant excitation~\cite{JayakumarPRL2013,Muller2014}. The optical excitation is performed with a Ti:Sapphire pulsed laser with $\SI{80}{\mega \hertz}$ repetition rate. A 4f pulse shaper with an adjustable slit on its Fourier plane is used to select the laser energy at half the ground-to-biexciton state transition energy, enabling two-photon absorption, and at a bandwidth of $\SI{1060}{\micro \electronvolt}$, to minimize state mixing induced by laser-induced level shifts~\cite{BassetPRL2023}. The power of the laser pulses is set at $\pi$-pulse of the Rabi oscillations, that is in the condition which maximizes brightness. To mitigate blinking, the QD charge environment is stabilized using an uncollimated halogen lamp with a blackbody spectrum~\cite{NguyenPRB2013}. The QD emits polarization-entangled photon pairs via the biexciton-exciton cascade. The two-qubit density matrix that describes the polarization state of the two correlated photons depends on a few parameters (see also Appendix~\ref{app:rotationPaulibasis}), including the fine structure splitting, that is the energy splitting between the two intermediate bright-exciton states of the radiative cascade. The effect of the fine structure splitting is to introduce a phase in the entangled state that depends on the randomly distributed emission time of the exciton-to-ground state transition~\cite{Hudson2007PRL,RotaIEEE2020}. If the acquisition is not time gated, this results in a mixed density matrix with a dominant term due to the $\ket{\phi^{+}} = 1/ \sqrt{2} (\ket{HH} + \ket{VV})$ state and an additional contribution from the $\ket{\phi^{-}} = 1/ \sqrt{2} (\ket{HH} - \ket{VV})$ state, whose relative weight depends on the ratio between fine structure splitting and exciton recombination lifetime.

The light emitted by the QD is collected by the same objective used for optical excitation and separated from the excitation path using a 90:10 beam splitter. The backscattered laser light is suppressed using tunable volume Bragg gratings with a spectral bandwidth of $\SI{0.4}{\nano \meter}$. A second set of volume Bragg gratings, operated in reflection mode, is employed to spectrally separate the emission from the two optical transitions that are at $\SI{793.55}{\nano \meter}$ and $\SI{795.39}{\nano \meter}$. The quantum tomography setup consists of two sets of a half-wave plate and a quarter wave plate (for state rotation), a polarizing beam splitter (for state projection), and two silicon avalanche photodiodes. The quarter wave plate rotated by the additional angular offset $\Delta$ is in the analyzer of the photons from the biexciton-to-exciton transition. The detectors have a time jitter of approximately $\SI{400}{\pico \second}$ (FWHM) and a detection efficiency of 46\% including receptacle losses. The single-photon events are recorded using a time-to-digital converter from Swabian Instruments with a resolution of $\SI{10}{\pico \second}$ (rms). Each two-qubit quantum state tomography is performed with two detectors per qubit in nine different polarization settings, amounting to a total of 36 coincidence measurements~\cite{AltepeterPRL2005}. The acquisition time for each polarization setting is $\SI{30}{\second}$ at a coincidence rate of approximately 27 kcps.

\subsection{Measurement data evaluation}
We provide details on the evaluation of the measurement data shown in the main text to allow the interested reader to reconstruct the results from the given measurement data. We use two source codes, one to determine an array of coincidence events from the raw detector clicks and the other one to evaluate the estimators and their distance.

\begin{table}[b]
\caption{\label{tab:tablestat} Statistical properties for non-Gaussian histograms
}
\begin{ruledtabular}
\begin{tabular}{ccr}
\textrm{$\Delta (\SI{}{\degree})$\footnote{we consider the lowest purity quantum state ($\SI{13.9}{\micro \electronvolt}$ fine structure splitting) for all angles}}&
mean $\mu$ in $10^{-3}$ \footnote{we obtain twice values of the order of $10^{-16}$ which is far below the detection resolution, thus we set the value equal to 0} &
standard deviation $\sigma$ in $10^{-3}$\\
\colrule
 50 & 0 & 1.5\\
 60 & 5 & 5\\
 130 & 0 & 0.5\\
\end{tabular}
\end{ruledtabular}
\end{table}

\begin{figure*}[ht]
    \centering
    \includegraphics{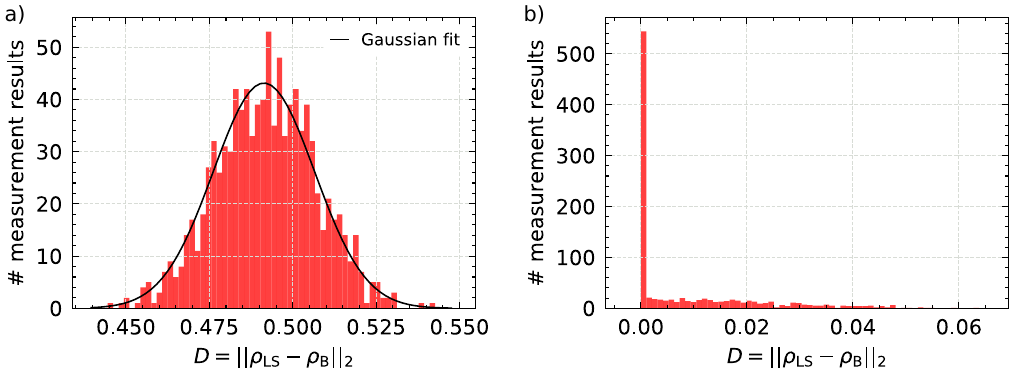}
\caption{a) The histogram summarizes the determined distance $D(\Delta=\SI{100}{\degree})$ for $1000$ sub-experiments for the state with the largest underlying state purity value ($\SI{0.6}{\micro \electronvolt}$ fine structure splitting). b) We observe for the state with the lowest purity that the resulting histogram for the distance $D(\Delta=\SI{60}{\degree})$ exhibits a non-Gaussian statistical behavior. The reason for the enhancement of zero-distance bin is that the nonphysical state lies inside the physical state space, because the purity is so low that even with systematic error the nonphysical estimator is a physical state. }
    \label{fig:Histograms}
\end{figure*}

For the evaluation of the estimator distance, we divide the total measured coincidence counts for each basis setting in $N_e=1000$ sub experiments consisting of $N_{bs} = 400$ coincidence events each. We split the array of measurement data in $N_{bs}=400$ blocks of size $N_e=1000$, whereas we assign each index in one block to one coincidence event of a sub-experiment. Subsequently, we evaluate for each sub-experiment the quantities of interest and bin the outcomes in a histogram. 

Fig.~\ref{fig:Histograms}\textcolor{blue}{a} and Fig.~\ref{fig:Histograms}\textcolor{blue}{b} show the histograms with the results for the $D(\Delta=\SI{100}{\degree})$ and $D(\Delta=\SI{60}{\degree})$ measurement points of the states with the largest and smallest underlying purity, respectively. The resulting histogram of the high purity quantum state exhibits a Gaussian statistics as seen in Fig.~\ref{fig:Histograms}\textcolor{blue}{a}, and thus, we fit a Gaussian distribution over the histogram. We use the mean value and the standard deviation of the Gaussian distribution as the measurement result and its error in Fig.~\textcolor{blue}{2c} of the main text, respectively. 

For the low purity results we observe for some angles $\Delta$ an outcome distribution, which differs clearly from the Gaussian distribution, see Fig.~\ref{fig:Histograms}\textcolor{blue}{b}. We observe this outcome distributions because the purity is so low that most of the, though erroneous, unbiased estimates still result in a physical state, and thus, the zero distance bin is filled up with most of the measurement results. For these specific cases we do not fit a Gaussian distribution over the histogram. In particular, we sum over all histogram values beginning from $D=0$ until we reach a cumulative value of $68.2\%$ of the total measurement outcomes, and we define the mean value and its standard deviation as the central bin and half of the width of the interval containing $68.2\%$ of the data. The reason for taking $68.2\%$ is that it corresponds with the definition of the standard deviation for the Gaussian function. Table \ref{tab:tablestat} contains all mean and standard deviation values for the low purity state with the offsets $\Delta$ for which the histogram is non-Gaussian.

\subsubsection{Confidence regions for measurement data} \label{app:confidence_regions_details}
\begin{figure}
    \centering
    \includegraphics{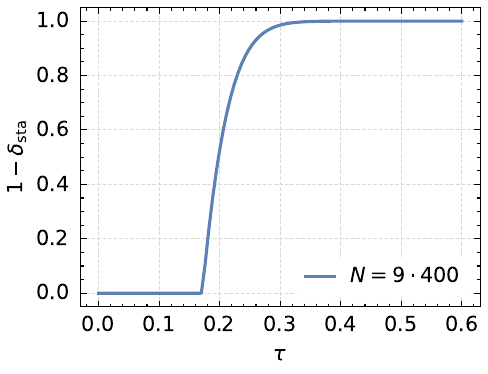}
    \caption{The confidence level $1-\delta_{\text{sta}}$ gives the probability that the distance between the estimators $D$ exceeds $\tau$ due to a systematic error for the experiment in the main text. We use Eq.~\textcolor{blue}{4} from the main text with the parameters $n=2$ and $N=9\cdot 400$ to obtain the confidence level as a function of $\tau$. We reach a confidence level of $90\%$ for a value of $\tau=0.25$.  }
    \label{fig:confidence_region}
\end{figure}

In this section, we analyze the confidence region for the experiment described in the main text. The confidence level is defined as $1-\delta_{\text{sta}}$, where $\delta_{\text{sta}}$ represents the probability that the distance $D$ between the biased and unbiased estimators exceeds $\tau$ due to a systematic error. This probability is provided by Eq.~\textcolor{blue}{4} in the main text. In our experiment, we focus on a two-qubit state, meaning $n=2$, and we perform a total of $N=9\cdot 400$ measurements. Fig.~\ref{fig:confidence_region} illustrates the confidence region as a function of $\tau$ from which we see that for a distance $\tau=0.25$ we get already a confidence level of $90\%$. Note that the bounds coming from the Bernstein inequality can get larger than $1$, which results in probabilities out of the interval $[0,1]$ and has been reported in the literature~\cite{Troop2015Bounds,GrossPRL2010Sensing}. Therefore, we adjust the probabilities in Fig.~\ref{fig:confidence_region} when we observe values of $\tau$ below $0$.

\section{Quantum state tomography of qubits implemented as the polarization of photons} \label{sec:app:Optics}
Here, we review concisely how to implement local, single-qubit Pauli measurements of qubits, which are realized by the polarization of a photon. In particular, we discuss the how to implement single-qubit unitary operations with wave plates, and further, we discuss how Pauli measurements are implemented in those photonic systems. 

\subsection{Single-qubit unitary operations}
The implementation of unitary rotations of the photon's polarization is crucial for the realization of local, single-qubit Pauli measurements. In our work we use quarter- and half-wave plates to perform unitary rotations on the photon's polarization, thus we summarize briefly the matrix formalism to treat wave plates, following the book of Pedrotti \ea~\cite{pedrotti_triple_2007}. A wave plate is an optical component (often cutout from a birefringent crystal), which causes an additional phase shifts for both the horizontal $\ket{H}$ and vertical $\ket{V}$ component of the photon's polarization. The most general form of a wave plate $\operatorname{W}$ is given by the following matrix,
\begin{equation}
\operatorname{W}=\left(\begin{array}{ll}
e^{\mathrm{i} \varepsilon_{x}} & 0 \\
0 & e^{\mathrm{i} \varepsilon_{y}}
\end{array}\right)=e^{\mathrm{i} \varepsilon_{H}}\left(\begin{array}{ll}
1 & 0 \\
0 & e^{\mathrm{i} (\varepsilon_{V}-\varepsilon_{H})}
\end{array}\right), \label{eq:app:generalPhaseRetarder}
\end{equation}
where $\varepsilon_{H}$ and $\varepsilon_{V}$ represent the phase change shift of the $H$ and $V$ component of the polarization. A quarter wave plate with fast axis vertically aligned, i.e. perpendicular to both the optical table and the photon's propagation direction, is obtained from Eq.~(\ref{eq:app:generalPhaseRetarder}) for phase shift of $\pi/4$ and $-\pi/4$ for the horizontal and vertical polarization component:
\begin{align}
    \operatorname{U}_{\mathrm{QWP,FAV}}=e^{\mathrm{i} \pi / 4}\left(\begin{array}{cc}
1 & 0 \\
0 & -\mathrm{i}
\end{array}\right). \label{eq:app:QWP_VFA}
\end{align}
Note that the matrix in Eq.~(\ref{eq:app:QWP_VFA}) is the $\operatorname{S}^{\dagger}$ gate. If the fast axis of the quarter wave plate is oriented horizontally, i.e., parallel to the optical table, the representing matrix differs from Eq.~(\ref{eq:app:QWP_VFA}) in terms of the sign of the relative and absolute phase,
\begin{equation} 
\operatorname{U}_{\mathrm{QWP,FAH}}=e^{-\mathrm{i} \pi / 4}\left(\begin{array}{ll}
1 & 0 \\
0 & \mathrm{i}
\end{array}\right). \label{eq:QWP_HFA}
\end{equation}
The matrix representation of a half-wave plate with fast axis oriented vertical is given as follows,
\begin{align}
 \operatorname{U}_{\mathrm{HWP,FAV}}=e^{\mathrm{i} \pi / 2}\left(\begin{array}{cc}
1 & 0 \\
0 & -1
\end{array}\right). \label{eq:app:UHWP}
\end{align}
A half-wave plate with fast axis oriented horizontal is described by the same matrix as in Eq.~(\ref{eq:app:UHWP}), merely the sign of the absolute phase term is opposite. Note that the matrix, Eq.~(\ref{eq:app:UHWP}), corresponds to the $\operatorname{Z}$ gate.

\subsection{Single-qubit Pauli basis measurements}
The photonic system we consider utilizes the polarization of a photon to implement the basis states of a qubit. Therefore, a computational basis measurement ($\PauliZ$ measurement) is implemented by observing clicks on a single-photon detector on each exit of a polarizing beam splitter (PBS). The remaining Pauli measurements are performed by inserting a quarter and half-wave plate prior to the PBS, see Fig.~\textcolor{blue}{2b} in the main text for a visualization of the measurement setup. The fast axis rotation angles $\psi_{k}$ and $\chi_{k}$ for the quarter and half-wave plate are determined from the expectation values of the $k$-th Pauli measurement setting as follows:
\begin{widetext}
\begin{align}
    \trace \left(\sigma_k \rho \right)  = \trace \left( \operatorname{R^{\text{rot}}}\left(\chi_{k}\right)  \PauliZ \operatorname{R^{\text{rot}}}\left(-\chi_{k}\right) \operatorname{R^{\text{rot}}}\left(\psi_{k}\right) \operatorname{S} \operatorname{R^{\text{rot}}}\left(-\psi_{k}\right)  \PauliZ    \operatorname{R^{\text{rot}}}\left(\psi_{k}\right) \operatorname{S}^{\dagger} \operatorname{R^{\text{rot}}}\left(-\psi_{k}\right) \operatorname{R^{\text{rot}}} \left(\chi_{k}\right) \PauliZ^{\dagger} \operatorname{R^{\text{rot}}}\left(-\chi_{k}\right) \rho \right),\label{eq:PauliBasiswave plates}
\end{align}
\end{widetext}
where $k\in \{1,\dots N_{\text{b}}\}$ is one of the $N_{\text{b}}$ measurement basis settings, and $\operatorname{R^{\text{rot}}}$ is the rotation matrix:
\begin{align}
\operatorname{R^{\text{rot}}}(\theta)=\left(
\begin{array}{cc}
 \cos (\theta ) & -\sin (\theta ) \\
 \sin (\theta ) & \cos (\theta ) \\
\end{array}
\right). \label{eq:DefinitionRotMatrix}
\end{align}
We summarize the rotation angles for the Pauli basis measurements in Table~\ref{tab:table1}. 
\begin{table}[b]
\caption{\label{tab:table1}Pauli measurement angles }
\begin{ruledtabular}
\begin{tabular}{ccc}
\text{Pauli basis}&$\psi$&$\chi$\\
\hline
$\PauliX$&$0$&$\frac{\pi}{8}$\\ \hline
$\PauliY$&$\frac{\pi}{4}$&$0$\\ \hline
$\PauliZ$&$0$&$0$\\
\hline
\end{tabular}
\end{ruledtabular}
\end{table}

\section{Experimental errors with polarization-entangled photon Pairs\label{app:furtherErrors}}
In this section we provide a non-exhaustive list of possible errors in a quantum optics laboratory, where the photon's polarization defines the qubit and local, single-qubit Pauli basis measurements are performed with wave plates and beam splitters. For other qubit implementations~\cite{PRLBrendelTimeBin} and optical components, as for example electro-optic modulators~\cite{EOMSpie}, other error settings may occur. 

\subsection{Rotation of a single Pauli basis}\label{app:rotationPaulibasis}
In the main text, we present the single-qubit case with a misalignment of a single Pauli basis. Here, we simulate two-qubit quantum states of varying purity emitted by quantum dots. Ideally, the emitted quantum state from a quantum dot is the Bell state $\ket{\phi^{+}}$, but fine structure splitting and the finite lifetimes of the exciton lead to a state deviating from $|\phi^{+}\rangle$. We use the model proposed by Hudson \ea~\cite{Hudson2007PRL} to describe a plausible quantum state $\rho$ generated by a quantum dot:
\begin{equation}
\rho=\frac{1}{4}\left(\begin{array}{cccc}
1+\kappa g_{H, V}^{\prime(1)} & 0 & 0 & 2 \kappa g_{H, V}^{(1)} z^* \\
0 & 1-\kappa g_{H, V}^{\prime(1)} & 0 & 0 \\
0 & 0 & 1-\kappa g_{H, V}^{\prime(1)} & 0 \\
2 \kappa g_{H, V}^{(1)} z & 0 & 0 & 1+\kappa g_{H, V}^{\prime(1)}
\end{array}\right) 
\end{equation} with
\begin{align}
    &g_{H, V}^{\prime(1)}=1 / \left(1+\tau_1 / \tau_{S S}\right), \nonumber \\ &g_{H, V}^{(1)}=1 /\left(1+\tau_1 / \tau_{S S}+\tau_1 / \tau_{H V}\right), \nonumber \\
&z=\frac{1+\mathrm{i} \iota}{1+\iota^2} \quad \text {and} \quad \iota=\frac{g_{H, V}^{(1)} S \tau_1}{\hbar}.
\end{align}
In this simulation we set the lifetime of the exciton $\tau_1  = \SI{150}{\pico \second}$, and the fraction of photon pairs originating exclusively from the
quantum dot $\kappa = 0.02$. We consider spin scattering and cross dephasing negligible by setting $\tau_\mathrm{SS} = \SI{1}{\micro \second}$ and $\tau_{\text{HV}} = \SI{1}{\micro \second}$. We vary the fine structure splitting from $0$ to $\SI{10}{\micro \electronvolt}$, which results in quantum states with purity ranging from $0.92$ to $0.58$, as in the experiment reported in the main text. 

\begin{figure*}
    \centering
    \includegraphics{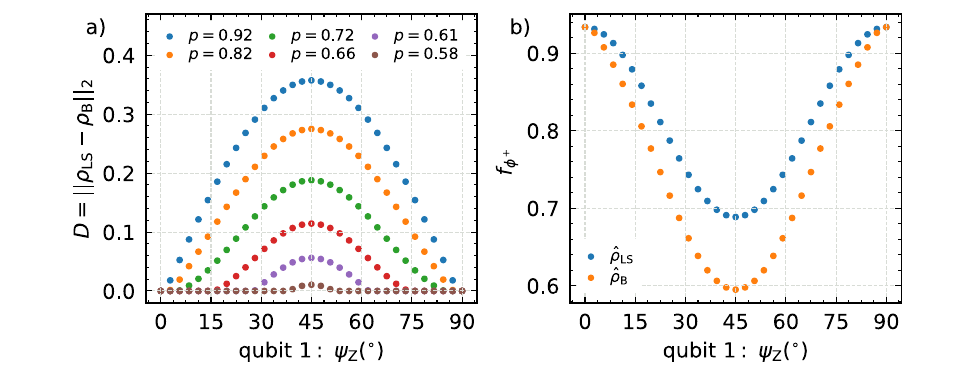}
    \caption{a) Distance $D$ as a function of rotation angle $\psi_{\text{Z}}$ of the quarter wave plate on qubit 1 for six different two-qubit state purities, ranging from $0.58$ to $0.92$. The rotation effects the basis terms: $\PauliZ \otimes \PauliX$, $\PauliZ \otimes \PauliY$ and $\PauliZ \otimes \PauliZ$. The maximum $D$ is found for an angle $\psi_{\text{Z}}=\SI{45}{\degree}$, where $\PauliY$ is measured for $\PauliZ$. The distance $D$ decreases for decreasing state purities. b) Fidelity w.r.t $|\phi^{+}\rangle$ evaluated for the least-squares estimator $\rhoLS$ (blue) and biased estimator $\rhoBiased$ (orange) depending on the rotation $\psi^{\text{Z}}$ of the quarter wave plate at qubit 1. The estimators are obtained from the measurement of the state with purity $0.92$. The fidelity values differ as function of the rotation angle and have a maximum at $\psi^{\text{Z}}=\SI{45}{\degree}$ where $\PauliY$ is measured instead of $\PauliZ$. }
    \label{fig:twoQubitAppendix}
\end{figure*}

We perform local, single-qubit Pauli measurements $\operatorname{P}_{\mathrm{qb1}} \otimes \operatorname{P}_{\mathrm{qb0}}$ with a fixed sequence of basis measurements: $\PauliX \otimes \PauliX$, $\PauliX \otimes \PauliY$,\dots $\PauliZ \otimes \PauliY$,$\PauliZ \otimes \PauliZ$ applied to qubit 1 and 0, respectively, such that the number of wave plate rotations is minimal. We assume that an error occurs during the adjustment of the $\PauliZ$ basis on qubit 1, which implies effects on the three basis measurements: $\PauliZ \otimes \PauliX$,$\PauliZ \otimes \PauliY$ and $\PauliZ \otimes \PauliZ$. Fig.~\ref{fig:twoQubitAppendix}\textcolor{blue}{a} shows the distance $D$ as function of the angle $\psi_{\text{Z}}$ on qubit 1 for quantum states with a purity ranging from $0.58$ to $0.92$. The maximum of the distance $D$ is found for the angle $\psi_{\text{Z}}=\SI{45}{\degree}$, this corresponds to the case of measuring $\PauliY$ instead of $\PauliZ$. Furthermore, we see that the magnitude of the distance $D$ decreases for decreasing values of the state purity, as discussed in the main text. 

Fig.~\ref{fig:twoQubitAppendix}\textcolor{blue}{b} shows the fidelity with respect to the Bell state $\ket{\phi^{+}}$ for the estimators $\rhoLS$ (blue) and $\rhoBiased$ (orange) as function of the angle $\psi_{\text{Z}}$ on qubit 1. Both types of estimators have been obtained from the measured quantum state with purity $0.92$. The resulting fidelity for the physical and nonphysical estimator differs from each other as function of $\psi_{\text{Z}}$. The maximal difference in fidelity is obtained for $\psi_{\text{Z}}=\SI{45}{\degree}$, where $\PauliY$ is measured instead of $\PauliZ$.

\subsection{Exchange of wave plate sequence}
Here, we consider the unintended exchange of the quarter and half-wave plate in the measurement setup of Fig.~\textcolor{blue}{2b}. We assume that the wave plates are perfectly aligned and the angles are adjusted as given in Table~\ref{tab:table1}. Mathematically, this error corresponds to exchanging the $\PauliZ (\PauliZ^{\dagger})$ and $\operatorname{S}^{\dagger} (\operatorname{S})$ between the rotation matrices $\operatorname{R}(\psi_{i})$ and $\operatorname{R}(\chi_{i})$ in Eq.~(\ref{eq:PauliBasiswave plates}), respectively. This error case has an effect on the implementation of the $\PauliX$ and $\PauliY$ basis,
\begin{align}
    &\sigma_{\mathrm{X, error}} = \frac{1}{2}\left(
\begin{array}{cc}
 1 & 1+\mathrm{i} \\
 1-\mathrm{i} & -1 \\
\end{array}
\right) = \PauliX - \PauliY + \PauliZ \nonumber \\ &\sigma_{\mathrm{Y, error}} = -\PauliZ = -\left(
\begin{array}{cc}
 1 & 0 \\
 0 & -1 \\
\end{array}
\right).
\end{align}
These errors caused on the $\PauliX$ and $\PauliY$ measurement are detectable under as discussed in Appendix~\ref{app:criticalPurity}.
 
\subsection{Misalignment of the quarter wave plate's fast axis}
Here, we discuss the cases of the unintended horizontal instead of vertical alignment of the quarter wave plate's fast axis, this exchange corresponds to the use of matrix Eq.~(\ref{eq:QWP_HFA}) instead of Eq.~(\ref{eq:app:QWP_VFA}) for the Pauli measurements. The same misalignment of the half-wave plate's fast axis causes no problems, because the resulting matrix Eq.~(\ref{eq:app:UHWP}) is up to a global phase factor the same for vertical and horizontal orientation of the fast axis. Mathematically, the horizontal orientation of the quarter wave plate corresponds to exchanging the $\operatorname{S}^{\dagger}$ and $ (\operatorname{S})$ gate in Eq.~(\ref{eq:PauliBasiswave plates}), which affects the $\PauliY$ basis measurement as follows,
\begin{align}
   \sigma_{\mathrm{Y, error}} = -\PauliY,
\end{align}
and this error corresponds performing a transposition of the single-qubit or a partial transpositon of the two-qubit quantum state, respectively. This error is not detectable, if we only have a single-qubit state or a two-qubit state product state. But as soon as the two-qubit state is entangled, we are able to detect this error as we have demonstrated in the main text and in Appendix~\ref{app:entanglementUseful}.

\subsection{Further errors}
For the sake of completeness, we mention briefly other possibilities for errors and provide references to works considering these errors in detail. 

The wave plates used in the Pauli basis measurement, can lead to errors effecting the matrices in Eqs.~(\ref{eq:app:QWP_VFA}, \ref{eq:QWP_HFA}, \ref{eq:app:UHWP}) of the quarter and half-wave plates, which may result in nonphysical quantum state estimates. West and Smith~\cite{ProceedingWestSmith1994} discuss polarization errors for zero-order wave plates caused by thickness mismatch, optical axis tilt and fast axis misalignment of the two crystals used in the wave plate. The polarization error caused by the light's incidence angle is discussed as well by West and Smith. Boulbry \ea~\cite{Boulbry:01} discuss the emerging of elliptical eigenpolarization modes caused by the misalignment of the optical axis a double crystal, achromatic, zero-order quarter wave plate.

Altepeter \ea~\cite{ALTEPETER2005105} provide an extensive discussion on errors of photonic polarization qubit measurements and how to correct them. They discuss errors caused by the crosstalk and absorption of the polarizing beam splitter as well as the effect of the efficiency mismatches of the two single-photon detectors used at the two beam splitter exits.

\bibliography{refs}

\end{document}